\documentclass[10pt,journal]{IEEEtran}

\usepackage{cite}
\ifCLASSINFOpdf
\else
  \usepackage[dvips]{graphicx}
  \DeclareGraphicsExtensions{.eps}
\fi
\usepackage[cmex10]{amsmath}
\usepackage{amsfonts,amssymb,amsthm,mdwmath}
\newtheorem{lemma}{Lemma}

\usepackage{array}

\ifCLASSOPTIONcompsoc
  \usepackage[caption=false,font=normalsize,labelfont=sf,textfont=sf]{subfig}
\else
  \usepackage[caption=false,font=footnotesize]{subfig}
\fi

\usepackage{stfloats}

\usepackage{url}
\usepackage{color}
\usepackage{acronym}
\begin{document}
\acrodef{PPP}[PPP]{Poisson Point Process}
\acrodef{CDF}[CDF]{Cummulative Distribution Function}
\acrodef{PDF}[PDF]{Probability Distribution Function}
\acrodef{RV}[RV]{Random Variable}
\acrodef{SIR}[SIR]{Signal-to-Interference Ratio}
\acrodef{SNR}[SNR]{Signal-to-Noise Ratio}
\acrodef{PLS}[PLS]{Physical layer security}
\acrodef{PGFL}[PGFL]{Probability Generating Functional}
\acrodef{CSI}[CSI]{channel state information}
\acrodef{i.i.d.}[i.i.d.]{independent and identically distributed}
\acrodef{w.r.t.}[w.r.t.]{with-respect-to}

\title{Boundaries as an Enhancement Technique for Physical Layer Security}
\author{Konstantinos Koufos and Carl P. Dettmann 
\thanks{K.~Koufos and C.P.~Dettmann are with the School of Mathematics, University of Bristol, BS8 1TW, Bristol, UK. \{K.Koufos, Carl.Dettmann\}@bristol.ac.uk} \protect \\ 
\thanks{This work was supported by the EPSRC grant number 
EP/N002458/1 for the project Spatially Embedded Networks.}}
\maketitle

\begin{abstract}
In this paper, we study the receiver performance with physical layer security in a Poisson field of interferers. We compare the performance in two deployment scenarios: (i) the receiver is located at the corner of a quadrant, (ii) the receiver is located in the infinite plane. When the channel state information (CSI) of the eavesdropper is not available at the transmitter, we calculate the probability of secure connectivity using the Wyner coding scheme, and we show that hiding the receiver at the corner is beneficial at high rates of the transmitted codewords and detrimental at low transmission rates. When the CSI is available, we show that the average secrecy capacity is higher when the receiver is located at the corner, even if the intensity of interferers in this case is four times higher than the intensity of interferers in the bulk. Therefore boundaries can also be used as a secrecy enhancement technique for high data rate applications.
\end{abstract}


\section{Introduction}
\label{sec:Introduction}
With the forecasted deployment of indoor ultra-dense wireless networks, it becomes important to develop models that consider the impact of boundaries in the performance analysis~\cite{Coon2012, Banani2015, Dettmann2015, Valenti2012, Heath2016, Koufos2016}. It is well-known that close to the boundary, the connection probability degrades due to  isolation~\cite{Coon2012, Dettmann2015}, but it improves in terms of interference~\cite{Banani2015, Koufos2016}. Analytical models considering finite deployment areas have so far been used to study spatial and temporal interference aspects~\cite{Banani2015, Valenti2012, Koufos2016}, optimize the base station density in cellular networks~\cite{Banani2015}, assess millimeter-wave network performance~\cite{Heath2016}, etc. 

\ac{PLS} without exchanging secret keys was first proposed by Wyner~\cite{Wyner1975}, and refers to the protection of information messages against eavesdropping with the aid of channel coding. \ac{PLS} would be well-suited for devices with light computational power, e.g., in certain types of sensor networks, where conventional security techniques fail~\cite{Trappe2015}. Nevertheless, the  impact of boundaries on connectivity and rate with \ac{PLS} has so far received limited attention. 

A great deal of research has adopted a type of random geometric graphs, known as the secrecy graph~\cite{Haenggi2008, Pinto2012, Pinto2010}, and studied the \ac{PDF} of the in- and out-connectivity degree with \ac{PLS}, the isolation probabilities, percolation threholds, etc. Another category of research considered the impact of interference on \ac{PLS}, and applied stochastic geometry to study the performance for the typical user in networks with infinite extent~\cite{Towsley2016,Vuppala2016,Heath2017, Deng2016}. In~\cite{Towsley2016}, the trade-off between the connection and the secrecy probabilities in cellular systems is studied, and in~\cite{Vuppala2016} it is shown that cluttered environments and blockage can be helpful in meeting secrecy constraints. In~\cite{Heath2017}, secure vehicle-to-vehicle communication is considered; a subset of antennas is used for beamforming towards the receiver, while the rest send jamming signals towards other directions. In~\cite{Deng2016}, relays forward the data between the sensors and the sinks, and their density is optimized for maximizing the average secrecy rate. 

\subsection{Related work $-$ Secrecy enhancement techniques}
In general, protecting the information messages against eavesdropping with \ac{PLS} comes along with a cost on the connection probability~\cite{Towsley2016} and the throughput~\cite{Zhou2011}. To mitigate the cost, secrecy enhancement techniques may be applied, especially when the density of eavesdroppers is high~\cite{Dhillon2017,Zheng2015}. When the \ac{CSI} of the intended receiver is known, eigen-beamforming can be used to maximize the \ac{SNR} of the intended channel~\cite{Zhou2011b}. Eigen-beamforming outperforms sectoring at the cost of knowing the \ac{CSI} instead of the direction~\cite{Zhou2011b}. Secrecy can be further enhanced when artificial noise is transmitted to the direction of the other sectors or to the null space of the intended channel~\cite{Zhang2013}. Combining artificial noise transmission with multi-antenna techniques is also considered in~\cite{Zheng2015}. In this study, the power levels of the information signal and the artificial noise are allocated to minimize the secrecy outage probability. The transmission of artificial noise works particularly well for secrecy enhancement, when the eavesdropper has fewer antennas than the transmitter, otherwise transmission of artificial fast fading achieves better secrecy because it prevents the eavesdropper from estimating the channel~\cite{Wang2015}. With single antenna equipment, it might be possible for the receivers to transmit jamming signals while receiving, provided they possess good self-interference cancellation mechanisms~\cite{Wang2016,Zheng2017}. Artificial noise and beamforming come  with a power and computational cost for the transmitter. Other alternatives for secrecy enhancement include multi-user scheduling~\cite{Zou2015, Lei2017} which  enhances the capacity of the main channel while leaving the capacity of the wiretap channel unaffected, and cooperative diversity which uses the best relay(s) in terms of secrecy capacity to forward the information messages~\cite{Zou2015}. Finally, when the transmitter can obtain some information about the location of the eavesdroppers, guard zones can be constructed; each transmitter will send confidential information when its guard zone is free from eavesdroppers, and the secrecy transmission capacity, especially under high security constraints, is enhanced~\cite{Zhou2011}.

The information theoretic approaches~\cite{Haenggi2008, Pinto2012, Pinto2010} and the analysis using stochastic geometry~\cite{Towsley2016,Vuppala2016,Heath2017,Deng2016,Zhou2011,Dhillon2017, Zheng2015,Zhou2011b,Zhang2013} assume that the locations of the transmitters and the eavesdroppers follow the uniform distribution in the infinite plane. To the best of our knowledge, the only available studies considering the impact of boundaries on secrecy performance are~\cite{Koufos2017,Liu2017}. The study in~\cite{Koufos2017} neglects the interference effects, and shows that the mean in- and out-connectivity degrees with \ac{PLS} in a quadrant are not necessarily equal, unlike in the infinite plane. The study in~\cite{Liu2017} considers a transmitter-receiver pair and a \ac{PPP} for the locations of  eavesdroppers inside an L-sided convex polygon. The secrecy rate is studied  for different $L$'s. Interference effects are neglected too.

\subsection{Related work $-$ Performance of wireless networks in confined areas}
The performance evaluation of wireless networks with irregular structure in the presence of interference has been mostly asymptotic, assuming a \ac{PPP} for the locations of base stations and users in a space with infinite extent~\cite{Andrews2011}. In practice, wireless networks are limited by physical boundaries once deployed indoors,   and they may also offer services over limited locations, e.g., public outdoor hotspots. Finite areas would naturally complicate the analysis because the notion of typical receiver is no longer valid; the performance becomes dependent on the location and the shape of the area. At the same time, the asymptotic assumption underestimates the performance for networks with low densities and also near the boundaries, where the interference would be naturally less~\cite{Banani2015}.

The moment generating function of interference due to a Binomial Point Process at the origin of a $d$-dimensional ball is derived in~\cite{Haenggi2007}. Over there, it is also shown that the \ac{PDF} of interference converges to Gaussian for a large number of interferers. The study in \cite{Zanella2009} extends the statistical analysis of interference for arbitrarily-shaped areas. When the point where the interference statistics are collected is located outside of the area generating the interference, e.g., primary-secondary system set-up, the moments of interference (also cross-moments) can be well-approximated using integration~\cite{Ruttik2011, Halim2014}.

The location-dependent property of outage probability over finite areas is also highlighted in~\cite{Guo2014} for ad hoc networks and in~\cite{Guo2017} for heterogeneous cellular networks. Finite deployment areas are often associated with a non-uniform \ac{PDF} of user location, as an attempt to model the impact of population density and/or mobility~\cite{Haenggi2014}. For a random waypoint mobility model, the mean interference at the origin is asymptotically twice the mean interference due to a uniform mobility model  because the users are concentrated towards the center of the area~\cite{Haenggi2014}. The temporal statistics of interference and outage become also location dependent, with higher correlation close to the boundary, where the degree of mobility is less~\cite{Koufos2018}.

\subsection{Motivation and list of contributions}
With boundaries, the interference field becomes nonhomogeneous. Therefore a natural question to ask is whether placing the receiver close to the boundary, where the interference is less, can enhance \ac{PLS}. Before looking at the impact of boundaries on the secrecy performance, let us consider the case, where the receiver and eavesdropper are deployed in the infinite plane (or in the bulk of the deployment area), and discuss the impact of interferer's intensity on the probability of secure connectivity, i.e., the joint event of successful decoding at the receiver and failure to decode at the eavesdropper~\cite{Are2011, Alouini2017}. We assume a single receiver and eavesdropper at fixed and known locations in a homogeneous Poisson field of interferers. The signal level over the main and the eavesdropper channels stay the same; it is only the interference level changing. When the intensity of interferers decreases, the interference level becomes less at the receiver and the eavesdropper. In that case, the probability of secure connectivity should decrease at low rates of the transmitted codewords (with reference to the Wyner encoding scheme), because the eavesdropper becomes capable of decoding low-rate transmissions almost surely. On the other hand, at high rates of the transmitted codewords, the probability of secure connectivity should increase because the performance is dominated by the connection probability of the receiver, which increases under a lower intensity of interferers. The above discussion gives an initial insight into the impact of boundaries on secure connectivity but it does not reveal the complete story. Placing the receiver close to the boundary  is not equivalent to placing the receiver in the bulk along with a reduction in the intensity of interferers. The boundary  introduces a trade-off which does not exist in the bulk and it is discussed next. 

Let us consider a quadrant, where the receiver is placed at the corner, i.e., at the point of minimum interference, and the eavesdropper along the side. The interference at the receiver and the eavesdropper is correlated because it is due to the same set of interferers~\cite{Haenggi2009}. We will show that the spatial correlation of interference is higher along the boundary than in the bulk, for the same distance separation between the receiver and the eavesdropper. Therefore placing the receiver at the corner is detrimental for \ac{PLS} because the reception conditions at the receiver and the eavesdropper become favorable at the same time. On the other hand, placing the receiver at the corner should benefit \ac{PLS} because the eavesdropper is exposed to higher interference than the receiver. The motivation of this paper is to study this interplay. 

The impact of interference correlation on the probability of secure connectivity in infinite cellular systems has been recently studied in~\cite{Alouini2017}. Over there it is shown that interference correlation plays a significant role in secrecy performance when the typical eavesdropper is located close to the typical user. In this paper, we consider an ad hoc type of system and compare the receiver performance at the corner and in the bulk of the deployment area considering both cases with known and unknown \ac{CSI} of the eavesdropper channel at the transmitter. We have in mind an indoor setting, e.g., industrial automation in a factory, smart home etc., where it is expected to have both high rate transmissions, e.g.,  video content using machine-to-machine technology, and/or low rate transmissions for exchanging measurement information and data fusion between low cost sensors. In the presence of eavesdroppers, we would like to identify whether it is beneficial to deploy the network elements near the boundaries or not and under which conditions on the transmission rate. The main findings are: 
\begin{itemize} 
\item When the \ac{CSI} of the eavesdropper is not available at the transmitter, it is beneficial to hide the receiver at the corner for high rates of the transmitted codewords because the performance is dominated by the connection probabilities of the receiver and the eavesdropper. At the corner, the receiver is exposed to lower interference as compared to an eavesdropper located along the boundary. 
\item  When the \ac{CSI} of the eavesdropper is not available at the transmitter, it is detrimental to hide the receiver at the corner for low rates of the transmitted codewords because an eavesdropper which is located along the boundary is also exposed to low interference thus, it can intercept the transmissions with high probability. 
\item When the transmitter can adapt the rate based on the instantaneous \ac{CSI}, the average capacity with \ac{PLS} is higher at the corner even if the intensity of interferers over there is four times higher than the intensity of interferers in the bulk. This means that the impact of higher interference at the eavesdropper than at the receiver dominates over the higher correlation of interference along the boundary than in the bulk. 
\end{itemize}

The remainder of this paper is organized as follows. In Section~\ref{sec:System}, we present the system model. In Section~\ref{sec:Connection}, we calculate the mean, the variance, the correlation coefficient of interference, and the connection probability of the receiver and the eavesdropper. In Section~\ref{sec:Secrecy}, we calculate the probability of secure connectivity assuming that the \ac{CSI} is not available at the transmitter. In Section~\ref{sec:Secrecy2}, we assume perfect knowledge of the \ac{CSI} and calculate the average secrecy capacity.  Section~\ref{sec:Secrecy} and Section~\ref{sec:Secrecy2} contain the main analysis of this paper and the comparison of the receiver performance at the boundary and in the bulk. In Section~\ref{sec:Conclusions}, we summarize the results of this paper and outline future work. 

\section{System model}
\label{sec:System}
\begin{figure}[!t]
 \centering
  \includegraphics[width=3.5in]{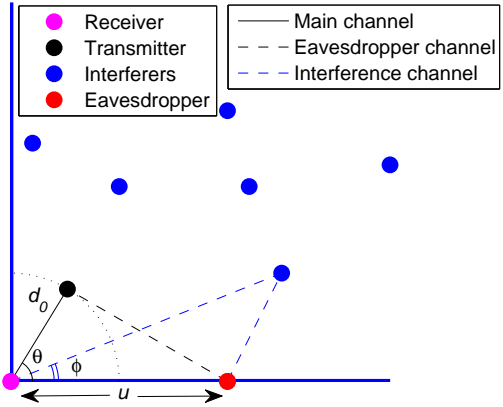}
 \caption{The geometry in which the receiver is located at the corner of the deployment area. The location of the eavesdropper is $\left(u,0\right)$.}
 \label{fig:illustration}
\end{figure}
We consider an ad hoc network where the locations of the transmitters follow  the \ac{PPP} with intensity $\lambda$, and each receiver is placed at a fixed link distance $d_0$ and a random angle $\theta$ from the associated transmitter, see Fig.~\ref{fig:illustration} for an illustration. The transmit power level is normalized to unity. We would like to assess the performance, i.e., connectivity and rate, with \ac{PLS}. In areas with boundaries, the performance is location-dependent. We consider two locations for the receiver: at the corner of a quadrant and in the bulk of the deployment area. In addition, we consider a single eavesdropper which is located at distance $u$ from the receiver. When the receiver is located at the corner, the eavesdropper is located at the boundary. The locations of the receiver and the eavesdropper are fixed and known. The origin of the coordinate system is set at the location of the receiver unless otherwise stated. The location of the eavesdropper is $\left(u,0\right)$. The transmitter, the receiver and the eavesdropper are equipped with a single antenna. The eavesdropper does not employ any advanced technique for intercepting the transmitter's message, e.g., successive interference cancellation, and the ad hoc network does not apply any secrecy enhancement technique, e.g., artificial noise. 

Considering just a single eavesdropper at an arbitrary location may seem overly simplistic, but it is used to get an insight on the comparison of secrecy performance with and without boundaries. Considering two-dimensional random locations for the eavesdroppers has been left as a future topic to study but the main conclusions of this paper are unlikely to change. After all, if we neglect eavesdroppers' collusion, a high (low) intensity of eavesdroppers means that the distance separation between the receiver and the most detrimental eavesdropper would be small (large), and the results of this paper are still applicable. For presentation brevity, we will also neglect the impact of interferers possibly located outside of the boundaries. Incorporating an additional interference field with a higher  propagation pathloss attenuation factor and/or penetration losses will increase the length of the expressions for the mean, the variance and the connection probability for the receiver located at the corner and for the eavesdropper along the boundary. Ignoring these interferers allows us to relate the statistics of interference for the receiver located at the corner and in the bulk in a simple manner. This facilitates the presentation of the proofs of lemmas, while the methodology and the conclusion of this paper will not change. In a practical system, one may also argue that the effect of interferers deployed outside of the boundaries would be negligible in case  millimeter wave propagation frequency is considered.

When the receiver is located at the corner, the location of the transmitter associated to it, hereafter the transmitter, is  $\left(d_0\cos\theta,d_0\sin\theta\right)$, where the \ac{RV} $\theta$ follows the uniform distribution in $\left[0,\frac{\pi}{2}\right]$, thus   $f_{\Theta}\!\left(\theta\right)\!=\!\frac{2}{\pi}$. In the bulk, the location of the transmitter should follow the uniform distribution in $\left[0,2\pi\right]$. Nevertheless, we would like to compare the performance at the two locations on a fair basis. In order to do that, the distribution of signal level over the eavesdropper channel should stay the same. Therefore we constrain the location of the transmitter over $\left[0,\frac{\pi}{2}\right]$ in both cases. We denote by $Z$ the \ac{RV} describing the distance-based propagation pathloss over the eavesdropper channel, $Z\!=\!g\left(\|d_0 e^{j\Theta} \!-\! u \|\right)$, where $g\left(r\right)\!=\!\min\left(1,r^{-\eta}\right)$ is the distance-based propagation pathloss function and $\eta\!>\!2$ is the  pathloss exponent. The \ac{PDF} $f_Z\!\left(z\right)$ is derived in the Appendix.

Due to the Slivnyak's Theorem, the locations of the transmitters generating interference to the receiver and the eavesdropper, hereafter the interferers (or the users), follow a \ac{PPP} with intensity $\lambda$. Their transmission probability is $\xi$. For a high intensity of active users $\lambda\xi$, the impact of noise can be ignored in the performance assessment. The fast fading $h$ over all channels, i.e., main channel, eavesdropper channel and interfering channels is \ac{i.i.d.} following the exponential \ac{PDF} with unit mean $\mathbb{E}\left\{h\right\}\!=\!1$. The assumption of independent fast fading between the receiver and the eavesdropper should be valid for distances $u$ larger than half the wavelength. We assume that the considered distances meet this constraint. 

In order to assess the performance with secrecy, we follow the Wyner encoding scheme~\cite{Wyner1975}, where the rate of transmitted codewords is $R_t$, and the rate of confidential messages is $R_s$. 
Let us denote by $\gamma_{\text{x,r}}$ the \ac{RV} describing the instantaneous \ac{SIR} at the receiver, where $\text{x}\!\in\!\left\{\text{bu,co}\right\}$ indicates the reveiver location in the bulk or at the corner. The connection probability of the receiver can be calculated as  $\mathbb{P}_{\text{x,r}}^{\text{c}} \!=\! \mathbb{P}\left\{ {\gamma_{\text{x,r}}} \!>\! \mu \right\}$, where $\mu \!=\! 2^{R_t}-1$. Similarly, let us denote by $\gamma_{\text{x,e}}\!\left(u\right)$ the \ac{RV} describing the \ac{SIR} at the eavesdropper, and by $\mathbb{P}_{\text{x,e}}^{\text{c}}\!\left(u\right)\!=\! \mathbb{P}\left\{ {\gamma_{\text{x,e}}\!\left(u\right)} \!>\! \sigma \right\}$ the probability that the eavesdropper succeeds to decode the transmitter's message. According to the Wyner scheme, $\sigma\!=\! 2^{R_e}-1$, where the rate $R_e\!=\! R_t\!-\! R_s$ reflects the rate cost to secure the message against the eavesdropper. For a positive secrecy rate $R_s\!\leq\! R_t$, it is required that $\mu\!\geq\!\sigma$. When the \ac{CSI} of the main and the eavesdropper channels is not available at the transmitter, the rates $R_t,R_s$ are kept fixed. A pair of rates $\left(R_t,R_s\right)$ can be associated with a probability of secure connectivity, $\mathbb{P}_{\text{x}}^{\text{sc}}$,  which can be expressed as the joint event~\cite{Are2011, Alouini2017} 
\begin{equation}
\label{eq:Psc0}
\mathbb{P}_{\text{x}}^{\text{sc}}\!\left(u\right)\!=\!\mathbb{P}\left( \gamma_{\text{x,r}}>\mu ,\gamma_{\text{x,e}}\!\left(u\right)<\sigma \right).
\end{equation}

When the \ac{CSI} at the receiver and the eavesdropper is perfectly known, the transmitter can adapt the transmission rate equal to $\max\!\left\{0,\log\left(\frac{1+\gamma_{\text{x,r}}}{1+\gamma_{\text{x,e}}}\right)\right\}$, and the performance is described in terms of average secrecy capacity~\cite{Barros2008, Wang2014}.  
\begin{equation}
\label{eq:Rate0}
\overline{C}_{\text{x}}^{\text{sc}}\!\!\left(u\right) = \!\! \int_0^\infty\! \!\!\int_0^{\gamma_{\text{x,r}}}\! \!\!\! \log_2\!\left(\frac{1+\gamma_{\text{x,r}}}{1+\gamma_{\text{x,e}}}\right)  \!f_{\text{r,e}}\!\left(\gamma_{\text{x,r}},\gamma_{\text{x,e}}\right) {\rm d}\gamma_{\text{x,e}} {\rm d}\gamma_{\text{x,r}}, 
\end{equation}
where $f_{\text{r,e}}\!\left(\gamma_{\text{x,r}},\gamma_{\text{x,e}}\right)$ is the joint \ac{PDF} of the \ac{SIR} at the receiver and the eavesdropper. 

Performance studies of \ac{PLS} with \ac{CSI} imperfections due to estimation errors at the receiver and/or limited feedback can be found in~\cite{He2013,Alouini2016} and references therein. Studying the impact of imperfections on the performance comparison with and without boundaries is a topic for future work.

While studying the performance with secrecy, we will need the mean and the variance of interference at the receiver and the eavesdropper, the correlation of interference between the two locations, and the connection probabilities. These quantities are calculated in the next section. 

\section{Interference and connection probability}
\label{sec:Connection}
In the bulk, the mean and the variance of interference are independent of the location. Therefore it suffices to calculate the moments of interference at the receiver 
\begin{equation}
\label{eq:Bulk}
\begin{array}{ccl}
\mathbb{E}\left\{\mathcal{I}_{\text{bu,r}}\right\} \!\!\!\!\!\!&=&\!\!\!\!\!\! \lambda\xi \int_0^\infty\int_0^{2\pi} g\left(r\right) r {\rm d}\phi \, {\rm d}r \stackrel{(a)}{=} \frac{\lambda\xi \eta \pi}{\eta-2}.  \\
\mathbb{V}{\text{ar}}\left\{\mathcal{I}_{\text{bu,r}}\right\} \!\!\!\!\!\!& = &\!\!\!\!\!\! 2 \lambda\xi \int_0^\infty\int_0^{2 \pi} g^2\!\left(r\right) r {\rm d}\phi\, {\rm d}r \stackrel{(b)}{=} \frac{2\lambda\xi\eta\pi}{\eta-1},  
\end{array}
\end{equation}
where $(a)$ and $(b)$ follow after taking into account the piecewise nature of the propagation pathloss function $g\!\left(\cdot\right)$, and the factor $2$ in the calculation of the variance comes from the second moment of a unit-mean exponential \ac{RV},  $\mathbb{E}\left\{h^2\right\}\!=\!2$.  

The mean and the variance of interference at the corner of a quadrant can be calculated after scaling the respective statistics in the bulk, see equation~\eqref{eq:Bulk}, by $\frac{1}{4}$, i.e., $\mathbb{E}\left\{\mathcal{I}_{\text{co,r}}\right\}\!=\!\frac{1}{4}\mathbb{E}\left\{\mathcal{I}_{\text{bu,r}}\right\}$ and $\mathbb{V}{\text{ar}}\left\{\mathcal{I}_{\text{co,r}}\right\}\!=\!\frac{1}{4}\mathbb{V}{\text{ar}}\left\{\mathcal{I}_{\text{bu,r}}\right\}$. In addition, the mean and the variance of the interference at the eavesdropper located at the boundary and at distance $u$ from the corner become easier to calculate after shifting the origin to $\left(u,0\right)$.  
\begin{equation}
\label{eq:EIVarI}
\begin{array}{ccl}
\mathbb{E}\left\{\mathcal{I}_{\text{co,e}}\!\left(u\right)\right\} \!\!\!\!\!\!&=&\!\!\!\!\!\! \lambda\xi \int_0^\infty\int_0^{\phi_{\text{co}}\!\left(u,r\right)} g\left(r\right) r {\rm d}\phi \, {\rm d}r.  \\
\mathbb{V}{\text{ar}}\left\{\mathcal{I}_{\text{co,e}}\!\left(u\right)\right\} \!\!\!\!\!\!& = &\!\!\!\!\!\! 2 \lambda\xi \int_0^\infty\int_0^{\phi_{\text{co}}\!\left(u,r\right)} g^2\!\left(r\right) r {\rm d}\phi\, {\rm d}r, 
\end{array}
\end{equation}
where $\phi_{\text{co}}\!\left(u,r\right)\!=\!\pi$ for $r\!\leq\!u$, and  $\phi_{\text{co}}\!\left(u,r\right)\!=\!\pi-\arccos\left(\frac{u}{r}\right)$ for $r\!>\!u$.

After differentiating equations~\eqref{eq:EIVarI} with respect to $u$ using the integral rule, one may show that the mean and the variance increase as we move away from the corner. Therefore an eavesdropper at the boundary is exposed to higher interference than the receiver at the corner. Due to the piecewise nature of the propagation pathloss function, we have to separate between two cases, $u\!\gtrless\!1$, in equation~\eqref{eq:EIVarI}, before expressing $\mathbb{E}\left\{\mathcal{I}_{\text{co,e}}\!\left(u\right)\right\}$ and $\mathbb{V}{\text{ar}}\left\{\mathcal{I}_{\text{co,e}}\!\left(u\right)\right\}$ in semi-closed form. 
\begin{equation}
\label{eq:EIVarI2}
\begin{array}{lll}
\mathbb{E}\!\left\{\!\mathcal{I}_{\text{co,e}}\!\left(u\right)\!\right\} \!\!\!\!\!\!&\stackrel{u<1}{=}&\!\!\!\!\!\! \lambda\xi\! \Big( \!\pi\!\int_0^u \!\! r {\rm d}r \!+\! \int_u^1\!\!\left( \pi \!-\! \arccos\left(\! \frac{u}{r} \!\right) \!\right) \! r {\rm d}r +\! \\ 
& & \int_1^\infty \!\!\left( \pi \!-\! \arccos\left( \frac{u}{r} \right) \right) r^{1-\eta} {\rm d}r \! \Big) \\  
\!\!\!\!\!\!& = &\!\!\!\!\!\! \lambda\xi\frac{\left(\eta-2\right)u\sqrt{1-u^2}+\eta\left(\pi-\arccos\left(u\right)\right)} {2\left(\eta-2\right)} \, - \\ 
& & \frac{u\,\, {}_2\!F_{\!1}\!\left(\frac{1}{2}, \frac{\eta-1}{2};\frac{\eta+1}{2};u^2\right)}{\left(\eta-1\right)\left(\eta-2\right)} \\ 
\mathbb{E}\!\left\{\!\mathcal{I}_{\text{co,e}}\!\left(u\right)\!\right\} \!\!\!\!\!\!&\stackrel{u\geqslant 1}{=}&\!\!\!\!\!\! \lambda\xi\!\left(\frac{\pi}{2\left(\eta-2\right)} - \frac{\sqrt{\pi}u^{2-\eta}\Gamma\left(\frac{\eta-1}{2} \right)}{2\left(\eta-2\right)\Gamma\left(\eta/2\right)} \right) \\ 
\mathbb{V}{\text{ar}}\!\left\{\!\mathcal{I}_{\text{co,e}}\!\left(u\right)\!\right\} \!\!\!\!\!\!&\stackrel{u<1}{=}&\!\!\!\!\!\!  2\lambda\xi\! \Big( \frac{u\sqrt{1-u^2}}{2}  + \frac{\eta\left(\pi-\arccos\left(u\right)\right)}{2\left(\eta-1\right)} \, - \\ & &\!\!\!\!\!\! \frac{u}{2\left(\eta-1\right)\left(2\eta-1\right)} {}_2\!F_{\!1}\!\left(\eta \!-\! \frac{1}{2},\!\frac{1}{2};\!\eta\!+\!\frac{1}{2};\!u^2\right)\!\! \Big)  \\ 
\mathbb{V}{\text{ar}}\!\left\{\!\mathcal{I}_{\text{co,e}}\!\left(u\right)\!\right\} \!\!\!\!\!\!&\stackrel{u\geqslant 1}{=}&\!\!\!\!\!\! 2\lambda\xi \frac{2\pi \eta \Gamma\left(\eta\right) - u^{2-2\eta}\sqrt{\pi}\, \Gamma\left(\eta- \frac{1}{2}\right) }{4\Gamma\left(\eta\right)\left(\eta-1\right)},
\end{array}
\end{equation}
where ${}_2F_1$ is the Gaussian hypergeometric function~\cite[pp.~556]{Abramo}, and $\Gamma\!\left(x\right)\!=\!\int_0^\infty t^{x-1} e^{-t} {\rm d}t$ is the Gamma function. 

In order to calculate the covariance of interference between the receiver and the eavesdropper, one should keep in mind that the set of interferers for the two locations are fully correlated. The Pearson correlation coefficient takes the following form: 
\begin{equation}
\label{eq:rho}
\rho_{\text{x}}\!\left(u\right) \!=\! \frac{ \lambda \xi \int_0^\infty \!\! \int_0^{\phi_{\text{x}}} g\!\left( r \right) g\!\left( \|re^{j\phi}\!-\!u\| \right) r {\rm d}\phi \, {\rm d}r} {\sqrt{\mathbb{V}{\text{ar}}\left\{\mathcal{I}_{\text{x,e}}\!\left(u\right)\right\}} \sqrt{ \mathbb{V}{\text{ar}}\left\{\mathcal{I}_{\text{x,r}}\right\}}}, 
\end{equation}
where $\text{x}\!\!\in\!\!\left\{\text{co},\!\text{bu}\right\}$,  $\phi_{\text{bu}}\!=\!2\pi, \phi_{\text{co}}\!=\!\frac{\pi}{2}$, and the interference in the bulk is independent of location,  $\mathbb{V}{\text{ar}}\!\left\{\!\mathcal{I}_{\text{bu,e}}\!\left(u\right)\!\right\}\!=\!\mathbb{V}{\text{ar}}\!\left\{\!\mathcal{I}_{\text{bu,r}}\!\right\} \! \forall u$.

The correlation coefficient $\rho_{\text{x}}$ is independent of the user density $\lambda$ and user activity $\xi$. In addition, we have seen that the user activity is just a scaling factor in the calculation of the mean and variance.  Hereafter, we omit the activity probability from the expressions for brevity, and the user intensity $\lambda$ describes the intensity of users after thinning with $\xi$.  

\begin{figure}[!t]
 \centering
  \includegraphics[width=3.5in]{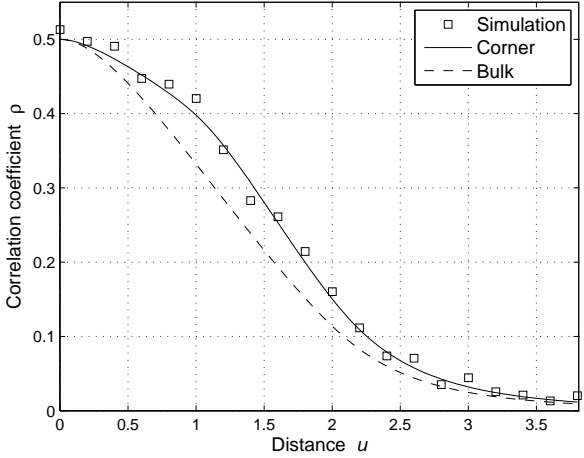}
 \caption{Spatial correlation coefficient of interference at distance $u$ from the receiver. The receiver is placed at the corner and in the bulk. In the numerator of equation~\eqref{eq:rho}, the integral is evaluated numerically. At the corner, equation~\eqref{eq:rho} is verified with simulations. Pathloss exponent $\eta\!=\!4$, and user density $\lambda\!=\!0.2$.}
 \label{fig:rho}
\end{figure}

In Fig.~\ref{fig:rho}, we depict the correlation coefficient at the corner and in the bulk with respect to the distance $u$. We see that placing the receiver at the corner increases the spatial correlation of interference for the same distance separation between the receiver and the eavesdropper. 

In order to calculate the connection probability of the receiver in the interference-limited regime, we need to evaluate the Laplace Transform of the interference $\mathbb{P}_{\text{x,r}}^{\text{c}} \!=\! \mathbb{E}\left\{e^{-s \mathcal{I}_{\text{x,r}}} \right\}$~\cite{Andrews2011}. Note that the impact of noise can be simply incorporated by scaling with a constant the Laplace Transform of the interference. Using the \ac{PGFL} of the \ac{PPP} we get 
\begin{IEEEeqnarray}{ccl}
\mathbb{P}_{\text{x,r}}^{\text{c}}   &=& \, \exp\!\!\left( \!-\lambda\!\int_0^\infty\!\!\! \int_0^{\phi_{\text{x}}}\!\! \frac{s g\!\left(r\right)}{1\!+\!sg\!\left(r\right)} r {\rm d}\phi \, {\rm d}r \right) \IEEEyesnumber\IEEEyessubnumber* \label{eq:Pca} \\ 
 &=& \, \exp\!\!\left(\!\!-\!\lambda \phi_{\text{x}}\!\!  \left(\!\! \frac{s}{2\left(1\!\!+\!\!s\right)} + \frac{s\, {}_2\!F_{\!1}\!\!\left(\! 1,\frac{\eta\!-\!2}{\eta};\frac{2\eta\!-\!2}{\eta};-\!s\!\right)} {\eta-2} \!\!\right)\!\!\! \right)\!\!, \label{eq:Pcb} 
\end{IEEEeqnarray}
where $s\!=\!\frac{\mu}{g\left(d_0\right)}$.

Let us assume for the moment that the location of the transmitter is fixed and known. In order to calculate the connection probability of the eavesdropper in the bulk, one should substitute $\phi_{\text{x}}\!=\!2\pi$, and $s_e\!=\!\frac{\sigma }{g\!\left(\|d_0 e^{j\theta}- u \|\right)}\!=\! \sigma z^{-1}$ instead of $s$ in equation~\eqref{eq:Pcb}. 
\begin{equation}
\label{eq:PcBulk}
\mathbb{P}_{\!\!\text{bu,e}}^{\text{c}}\!\!\left(u\right) \!=\! \exp\!\!\left(\!\!-\!2\pi\!\lambda\!\! \left(\!\! \frac{s_e}{2\left(1\!\!+\!\!s_e\right)} \!+\! \frac{s_e {}_2\!F_{\!1}\!\!\left(\! 1,\!\frac{\eta\!-\!2}{\eta};\!\frac{2\eta\!-\!2} {\eta};\!-s_e\!\right)} {\eta-2} \!\!\right)\!\!\! \right)\!\!\!.
\end{equation}
\begin{figure}[!t]
 \centering 
  \includegraphics[width=3.5in]{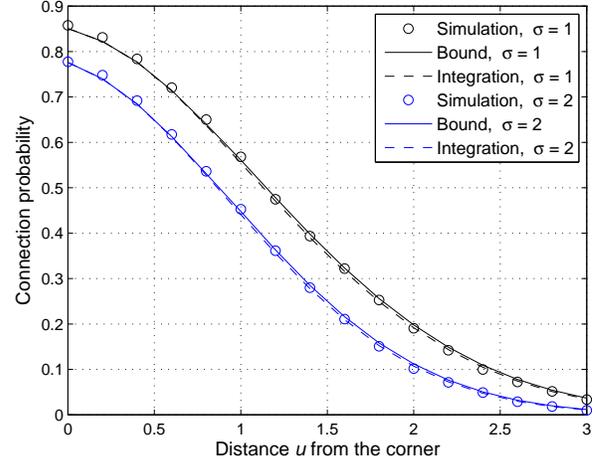}
 \caption{Connection probability for the eavesdropper \ac{w.r.t.} the distance $u$ from the corner. Pathloss exponent $\eta\!=\!4$, and user density $\lambda\!=\!0.2$. The transmitter is located at the boundary at $\left(0,1\right)$. The calculation uses equations~\eqref{eq:Pc1a} and~\eqref{eq:Pc1b} and the integrals $I_1, I_u$ are evaluated numerically. The bound uses the approximations in equations~\eqref{eq:Iua} and~\eqref{eq:Iub}.}
 \label{fig:Pr}
\end{figure}

When the eavesdropper is located at the boundary, one should substitute $s_e$ instead of $s$, and $\phi_{\text{co}}\!\left(u,r\right)$ instead of $\phi_{\text{x}}$ in equation~\eqref{eq:Pca}. After shifting the origin to $\left(u,0\right)$ and separating between $u\!\gtrless\! 1$ in the double integral in equation~\eqref{eq:Pca},  we get 
\begin{IEEEeqnarray}{ccl}
\label{eq:Pc1}
\mathbb{P}_{\!\!\text{co,e}}^{\text{c}}\!\left(u\right) &\stackrel{u<1}{=}& \!\!\exp\!\!\left(\!\!-\!\frac{\lambda s_e}{2\!\left(\!1\!\!+\!s_e\!\right)}\!\!\left( \!\pi \!\!+\! u\sqrt{\!1\!-\!u^2\!} \!-\! \arccos\!\left(u\right)\!\! \right) \!\!-\!\! \lambda I_{\!1}\!\! \right)  \IEEEyesnumber\IEEEyessubnumber* \label{eq:Pc1a} \\ 
\mathbb{P}_{\!\!\text{co,e}}^{\text{c}}\!\left(u\right)  &\stackrel{u\geqslant 1}{=}& \!\!\exp\!\!\Big(\!\!\!-\!\!\!\lambda\pi s_e\! \Big(\! \frac{1}{2\!\left(1\!+\!s_e\right)} \!+\! \frac{ {}_2F_1\!\!\left(1,\!\frac{\eta\!-\!2}{\eta};\!\frac{2\eta\!-\!2}{\eta};\!-\!s_e \right) \!\!} {\eta\!-\!2} -\!\! \IEEEnonumber \\ 
& & \! \frac{u^{2\!-\!\eta}}{\eta-2} \,\, {}_2F_1\!\! \Big( 1,\!\frac{\eta\!-\!2}{\eta};\!\frac{2\eta\!-\!2}{\eta};\!-\!\frac{s_e}{u^\eta} \!\Big)\Big) -\lambda I_u \!\Big)\!, \label{eq:Pc1b}
\end{IEEEeqnarray} 
where  $I_1\!=\! \int_1^\infty \left(\pi\!-\!\arccos\left(\frac{u}{r}\right)\right) \frac{s_e\, r }{s_e+r^\eta} {\rm d}r$, and $I_u\!=\! \int_u^\infty \left(\pi\!-\!\arccos\left(\frac{u}{r}\right)\right) \frac{s_e\, r }{s_e+r^\eta} {\rm d}r$. 

If we bound the inverse trigonometric function,  $\pi\!-\!\arccos\left(\frac{u}{r}\right) > \frac{\pi}{2}+\frac{u}{r},  \forall r \!\geq\! u$, we get a tight upper bound on the connection probability of the eavesdropper after substituting the following lower bound approximations in~\eqref{eq:Pc1}. 
\begin{IEEEeqnarray}{ccl}
I_1 \,\, &\gtrsim& \,\, \frac{u s_e^{1/\eta}}{ {\text{sinc}}\!\left(\pi/\eta\right)} -u\, {}_2F_1\left(1,\frac{1}{\eta};\frac{\eta+1}{\eta};-\frac{1}{s_e}\right) + \IEEEnonumber* \\ 
& & \,\,\,\,\,\,\, \frac{\pi s_e}{2\left(\eta-2\right)} \, {}_2F_1\left( 1,\frac{\eta-2}{\eta}; \frac{2\eta-2}{\eta};-s_e\right)  \IEEEyesnumber\IEEEyessubnumber* \label{eq:Iua} \\ 
I_u \,\, &\gtrsim& \,\, \frac{u s_e^{1/\eta}}{ {\text{sinc}}\!\left(\pi/\eta\right)} -u^2{}_2F_1\left(1,\frac{1}{\eta};\frac{\eta+1}{\eta};-\frac{u^\eta}{s_e}\right) + \IEEEnonumber \\ 
& & \,\,\,\,\,\,\, \frac{\pi s_e u^{2-\eta}}{2\left(\eta-2\right)} \, {}_2F_1\left( 1,\frac{\eta-2}{\eta};\frac{2\eta-2}{\eta};-\frac{s_e}{u^\eta}\right)\! , 
\label{eq:Iub}
\end{IEEEeqnarray} 
where ${\text{sinc}}\!\left(x\right)\!=\!\frac{\sin\left(x\right)}{x}$.

The tightness of the bound is illustrated in Fig.~\ref{fig:Pr}. The connection probability of the eavesdropper decreases rapidly along the boundary because the interference becomes higher over there, and at the same time the signal level over the eavesdropper channel decreases. The trend is similar when the location of the transmitter follows the uniform distribution. In that case, the connection probability can be calculated after integrating (numerically) equations~\eqref{eq:PcBulk}$-$\eqref{eq:Pc1b} over the \ac{PDF} of the signal level over the eavesdropper channel $f_Z\!\left(z\right)$. 

The connection probabilities for fixed and known transmitter's  location given in equations~\eqref{eq:PcBulk} and~\eqref{eq:Pc1b} would be of use in Section~\ref{sec:Secrecy} while approximating the probability of secure connectivity at high transmission rates $R_t$ and large distance separation $u$ between the receiver and the eavesdropper. For a large $u$, the correlation coefficient of interference may become negligible, see Fig.~\ref{fig:rho}, and the probability of secure connectivity can be approximated as the product of the connection probability of the receiver with the complementary of the connection probability of the eavesdropper. We will expand the connection probability of the eavesdropper for $\sigma\!\rightarrow\! 0/\sigma\!\rightarrow\!\infty$ to approximate the probability of secure connectivity for high/low secrecy rates $R_s$ under the assumption of uncorrelated interference. 

\section{Secure connectivity $-$ unknown \ac{CSI}}
\label{sec:Secrecy}
Using that the fading over the main and the eavesdropper channels is Rayleigh, the probability of secure connectivity in equation~\eqref{eq:Psc0} can be read as
\begin{equation}
\label{eq:Psc}
\begin{array}{ccl}
\mathbb{P}_{\text{x}}^{\text{sc}}\!\left(u\right) \!\!\!\!\!\!& = &\!\!\!\!\! \mathbb{E}\left\{ e^{-\!s\,\mathcal{I}_{\text{x,r}}} \left(1-e^{-\!s_e\mathcal{I}_{\text{x,e}}\left(u\right)}\right)\right\} = \mathbb{P}_{\text{x,r}}^{\text{c}} \!-\!  \mathcal{J}_{\text{x}}\!\left(u\right)\!, 
\end{array}
\end{equation}
where $\mathcal{J}_{\text{x}}\!\left(u\right) = \mathbb{E}\left\{ e^{-s\mathcal{I}_{\text{x,r}}-s_e\mathcal{I}_{\text{x,e}}\left(u\right)}\right\}$ is the joint connection probability of the receiver and the eavesdropper.  

In a recently published paper~\cite{Alouini2017}, the quantity $\mathcal{J}_{\text{x}}\!\left(u\right)$ has been calculated taking into account the fact that the interference at the receiver and the eavesdropper is correlated.  In order to take into account the correlation of interference in our problem setting, we condition on the location $\theta$ of the transmitter, and we  average over the fading states of the interfering channels at the receiver and the eavesdropper, as well as over the locations and activities of the interferers. After using the \ac{PGFL} of the \ac{PPP} and the fact that the fading samples in the interfering channels at the receiver and the eavesdropper are \ac{i.i.d.} unit-mean exponential \acp{RV} we get~\cite{Alouini2017}
\begin{eqnarray}
\label{eq:A}
\mathcal{J}_{\text{x}}\!\left(u\right) \!\!\!\!\!&=&\!\!\!\!\!\!\! \int_0^{\frac{\pi}{2}} \!\!\!\!\!\!  \exp\left(\!-\! \lambda\!\!\! \int\limits_{S_{\text{x}}}\!\! \left(1 \!-\! \frac{1}{1 \!+ \! s  g\left(r\right)} \frac{1}{1 \!+\! s_e g\!\left(d  \right)}\right) {\rm d}S \right) \!\!f_\Theta {\rm d}\theta \IEEEnonumber* \\
\!\!\!\!\!&=&\!\!\!\!\!\!\! \int\limits_Z \!\!\! \exp\!\!\left(\!\!\!-\! \lambda\!\!\! \int\limits_{S_{\text{x}}} \!\! \left(1 \!\!-\!\! \frac{1}{1 \!\!+ \! s  g\left(r\right)} \! \frac{1}{1 \!\!+\! \sigma z^{-\!1} \!g\!\left(d\right)}\right)\! {\rm d} S\! \right) \!\!f_Z {\rm d}z 
\end{eqnarray}
where $S_{\text{x}}$ is the infinite plane for $\text{x}\!=\! \text{bu}$ and the upper-right quadrant for $\text{x}\!=\! \text{co}$, ${\rm d}S \!=\! r {\rm d}r {\rm d}\phi$ is the integration element, $d \!=\! \|r e^{j\phi}-u\|$ is the distance between the integration element and the eavesdropper, $s\!=\!\frac{\mu}{g\left(d_0\right)}$, $s_e\!=\!\sigma z^{-\!1}$, and $z\!=\! g\!\left(\|d_0 e^{j\theta}- u \|\right)$ is the realization of the \ac{RV} $Z$ describing the distance-based pathloss over the eavesdropper channel. 
\begin{figure*}[!t]
 \centering
  \subfloat[$R_s\!\geq\! 0$]{\includegraphics[width=3.0in]{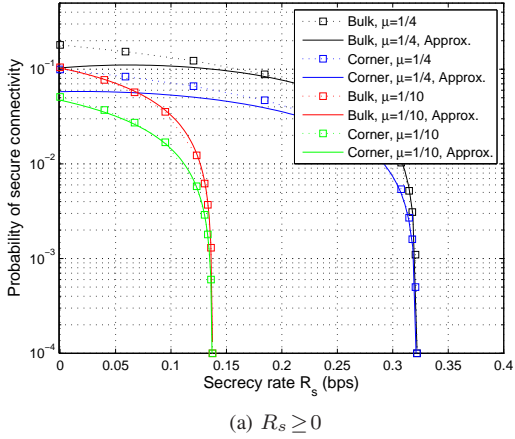} \label{fig:SmallTau}}
\hfil
  \subfloat[$R_s\!=\! 0$]{\includegraphics[width=3.0in]{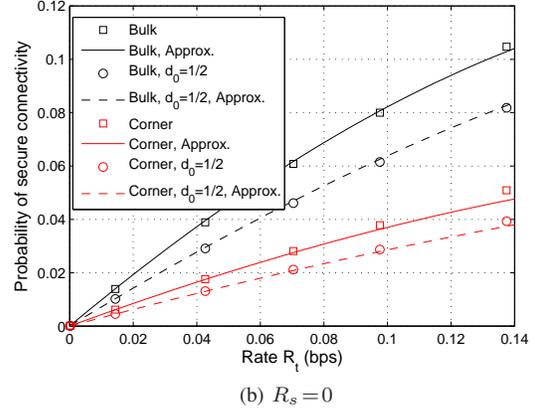}
 \label{fig:SmallTau0}}
 \caption{Validating the approximations for the probability of secure connectivity at low transmission rates $R_t$. (a) The approximation is given in~\eqref{eq:SmallTau}. The rate of the transmitter codewords is $R_t\!=\! \log_2\!\left(1+\mu\right)$. (b) The approximation is given in~\eqref{eq:SmallTau0}. The secrecy rate is $R_s\!=\! 0$, i.e., $\mu\!=\! \sigma$. In both figures, the exact probability is calculated numerically based on~\eqref{eq:Psc} and~\eqref{eq:A}. 
Pathloss exponent $\eta\!=\!4$, user density $\lambda\!=\!0.2$, $u\!=\! 1$, $d_0\!=\!1$ unless otherwise stated.}
\label{fig:NewSmallTau}
\end{figure*}
\begin{lemma}
\label{lemma:4}
For low transmission rates $R_t$, the probability of secure connectivity is higher in the bulk than at the corner. 
\begin{proof}
A low rate $R_t$ necessitates a low \ac{SIR} threshold $\mu$. After expanding around $s\!=\!0$ equation~\eqref{eq:Pca}, keeping up to the second order terms, we can approximate the connection probability for low rates $R_t$ as follows 
\begin{IEEEeqnarray}{lll}
\label{eq:Pct0}
\mathbb{P}_{\text{x,r}}^{\text{c}} &\approx& \exp\!\left(\!-\!\lambda \left( \int_{S_{\text{x}}}\left( s g\!\left(r\right) - s^2\! g^2\!\left(r\right) \right) {\rm d}S \right)\right) \IEEEnonumber* \\ 
&=&  \exp\!\left( -s \mathbb{E}\!\left\{ \mathcal{I}_{\text{x,r}}\right\} + \frac{s^2}{2}\mathbb{V}{\text{ar}}\!\left(\mathcal{I}_{\text{x,r}}\right) \right)  \\ 
 &\approx&  1 - s \mathbb{E}\!\left\{\mathcal{I}_{\text{x,r}}\right\} + \frac{s^2}{2} \left(\mathbb{V}{\text{ar}}\!\left(\mathcal{I}_{\text{x,r}}\right) + \mathbb{E}\!\left\{\mathcal{I}_{\text{x,r}}\right\}^2 \right). \IEEEyesnumber*
\end{IEEEeqnarray}

In order to approximate the quantity $\mathcal{J}_{\text{x}}\!\left(u\right)$ in equation~\eqref{eq:A} for low rates $R_t$, we expand around $\sigma  z^{-1}\!=\!0$ and $s\!=\! 0$, again keeping up to the second order terms. 
\begin{equation}
\label{eq:SmallTau00}
\begin{array}{lll}
\mathcal{J}_{\text{x}}\!\left(u\right) \!\!\!\!\!\!\! &\approx&\!\!\!\!\!\! \displaystyle \int_Z\!\! \exp\!\Bigg(\!\!-\!\lambda\!\!\int_{S_{\text{x}}}\!\!\!\Big(s g\!\left(r\right)\!+\! \sigma z^{-1} \!g\!\left(d\right) \!-\!\! s^2\!g^2\!\!\left(r\right) - \\ 
& & \sigma^2\!z^{-2}\!g^2\!\!\left(d\right)\!-\! s \sigma z^{-1}\!g\!\left(r\right) g\!\left(d\right)\!\Big) {\rm d}S\Bigg) \! f_Z \, {\rm d}z\\ 
\!\!\!\!\!\!\! &\approx&\!\!\!\!\!\!  1 \!\!-\!\! s \mathbb{E}\!\left\{\!\mathcal{I}_{\text{x,r}}\!\right\} \!\!-\!\!  \sigma \mathbb{E}\!\left\{\!\!Z^{-\!1}\!\right\}\!\!\mathbb{E}\!\left\{\!\mathcal{I}_{\text{x,e}}\!\left(u\right)\!\right\} \!\!+\!\! \frac{s^{\!2}}{2}\! \mathbb{V}{\text{ar}}\!\left(\!\mathcal{I}_{\text{x,r}}\right) \!+\! \\ 
& &  \!\!\! \frac{\sigma^2}{2}\mathbb{E}\!\left\{Z^{-2}\right\}\mathbb{V}{\text{ar}}\!\left(\mathcal{I}_{\text{x,e}}\!\left(u\right)\right) + \\   & &  
\!\!\!\rho_{\text{x}}\!\left(u\right)\!s \sigma\mathbb{E}\!\left\{Z^{-1}\right\} \!\!\sqrt{\mathbb{V}{\text{ar}}\!\left(\mathcal{I}_{\text{x,r}}\right) \!\mathbb{V}{\text{ar}}\!\left(\mathcal{I}_{\text{x,e}}\!\left(u\right)\right)}  +\! \\ 
& &  \!\!\! \frac{\sigma^2}{2}\mathbb{E}\!\left\{\!Z^{-\!1}\!\right\}^2\!\mathbb{E}\!\left\{\mathcal{I}_{\text{x,e}}\!\left(u\right)\right\}^2 \!\! +  \! \frac{s^2}{2}\mathbb{E}\!\left\{\mathcal{I}_{\text{x,r}}\right\}^2 \!+\! \\ 
& &  \!\!\! s \sigma\mathbb{E}\!\left\{Z^{-1}\right\}\rho_{\text{x}}\!\left(u\right)\sqrt{\mathbb{V}{\text{ar}}\!\left(\mathcal{I}_{\text{x,r}}\right) \!\mathbb{V}{\text{ar}}\!\left(\mathcal{I}_{\text{x,e}}\!\left(u\right)\right)}.
\end{array}
\end{equation}
\noindent
After subtracting equation~\eqref{eq:SmallTau00} from equation~\eqref{eq:Pct0}  we get
\begin{equation}
\label{eq:SmallTau}
\begin{array}{ccl}
\mathbb{P}_{\text{x}}^{\text{sc}}\!\left(u\right) \!\!\!\!&\approx&\!\!\!\! \sigma\mathbb{E}\!\left\{Z^{-1}\right\}\mathbb{E}\!\left\{\mathcal{I}_{\text{x,e}}\!\left(u\right)\right\} \!-\! \\ 
& &\!\!\!\!\!\!\!\!\!\!\!\!\!\!\!\!\!\!\!\!\! \frac{\sigma^2}{2}\!\! \left(\!\mathbb{E}\!\left\{\!Z^{-\!2}\!\right\}\!\!\mathbb{V}{\text{ar}}\!\left(\mathcal{I}_{\text{x,e}}\!\left(u\right)\right) \!+\! \mathbb{E}\!\left\{\!Z^{-\!1}\!\right\}^{\!2}\!\mathbb{E}\!\left\{\mathcal{I}_{\text{x,e}}\!\left(u\right)\right\}^{\!2}\right) \!-\!\! \\ 
& &  \!\!\!\!\!\!\!\!\!  2\rho_{\text{x}}\!\left(u\right) s \sigma \mathbb{E}\!\left\{Z^{-1}\right\} \sqrt{\mathbb{V}{\text{ar}}\!\left(\mathcal{I}_{\text{x,r}}\right) \!\mathbb{V}{\text{ar}}\!\left(\mathcal{I}_{\text{x,e}}\!\left(u\right)\right)}. 
\end{array}
\end{equation}

Recall that in the bulk the interference is independent of the location $u$, and the probability for secure connectivity can be simplified after substituting $\mathbb{E}\left\{\mathcal{I}_{\text{bu,r}}\right\}$ instead of $\mathbb{E}\!\left\{\mathcal{I}_{\text{bu,e}}\!\left(u\right)\right\}$ in equation~\eqref{eq:SmallTau}. In addition, due to the fact that $\mathbb{E}\!\left\{\mathcal{I}_{\text{co,e}}\!\left(u\right)\right\}\!\leq\!\lim\limits_{u\rightarrow\infty}\mathbb{E}\!\left\{\mathcal{I}_{\text{co,e}}\!\left(u\right)\right\} \!=\!\frac{1}{2} \mathbb{E}\!\left\{\mathcal{I}_{\text{bu,r}}\right\} \forall u$, the probability for secure connectivity at the corner, for low rates $R_t$, can be upper-bounded after substituting $\frac{1}{2} \mathbb{E}\!\left\{\mathcal{I}_{\text{bu,r}}\right\}$ instead of $\mathbb{E}\!\left\{\mathcal{I}_{\text{co,e}}\!\left(u\right)\right\}$ in the first-order term in equation~\eqref{eq:SmallTau}. Finally, we get that  $\lim_{\sigma\rightarrow 0, s \rightarrow 0 } \frac{\mathbb{P}_{\text{co}}^{\text{sc}}\left(u\right)}{\mathbb{P}_{\text{bu}}^{\text{sc}}\left(u\right)} \!=\! \frac{\mathbb{E}\!\left\{Z^{-1}\right\}\mathbb{E}\!\left\{\mathcal{I}_{\text{co,e}}\!\left(u\right)\right\}}{\mathbb{E}\!\left\{Z^{-1}\right\}\mathbb{E}\left\{\mathcal{I}_{\text{bu,e}}\!\left(u\right)\right\}} \!\leq \! \frac{\mathbb{E}\left\{\mathcal{I}_{\text{bu,r}}\right\}}{2\mathbb{E}\left\{\mathcal{I}_{\text{bu,r}}\right\}} \!=\! \frac{1}{2}$.  
\end{proof}
\end{lemma}

Lemma~\ref{lemma:4} can be intuitively explained as follows. At the boundary, the interference is low, thus both the receiver and the eavesdropper are capable of decoding low rate transmissions almost surely. Because of that, secure connectivity degrades. On the other hand, in the bulk, where the mean interference is at least twice than that at the boundary, there might be network instances where the eavesdropper may fail to decode a low rate transmission due to unfortunate interference conditions, and at the same time the receiver can successfully decode.

A case of particular interest is $R_s\!=\! 0$, or equivalently, $\mu\!=\!\sigma\!\triangleq\!\gamma$. In that case,  equation~\eqref{eq:SmallTau} is simplified to 
\begin{equation}
\label{eq:SmallTau0}
\begin{array}{ccl}
\mathbb{P}_{\text{x}}^{\text{sc}}\!\left(u\right) \!\!\!\!&\approx&\!\!\!\! \gamma\mathbb{E}\!\left\{Z^{-1}\right\}\mathbb{E}\!\left\{\mathcal{I}_{\text{x,e}}\!\left(u\right)\right\} - \\ 
& &\!\!\!\!\!\!\!\!\!\!\!\!\!\!\!\!\!\!\!\!\! \frac{\gamma^2}{2}  \Big(\mathbb{E}\!\left\{\!Z^{-\!2}\!\right\}\!\mathbb{V}{\text{ar}}\!\left(\mathcal{I}_{\text{x,e}}\!\left(u\right)\right) \!+\!    \mathbb{E}\!\left\{\!Z^{-\!1}\!\right\}^2\mathbb{E}\!\left\{\mathcal{I}_{\text{x,e}}\!\left(u\right)\right\}^2 \!\!+\! \\ 
& &  \!\!\!\!\!\!\!\!\! \frac{4\rho_{\text{x}}\!\left(u\right)}{g\!\left(d_0\right)}\mathbb{E}\!\left\{Z^{-1}\right\} \sqrt{\mathbb{V}{\text{ar}}\!\left(\mathcal{I}_{\text{x,r}}\right) \!\mathbb{V}{\text{ar}}\!\left(\mathcal{I}_{\text{x,e}}\!\left(u\right)\right)}\Big). 
\end{array}
\end{equation}

The accuracy of approximation~\eqref{eq:SmallTau} for the probability of secure connectivity at the corner and in the bulk is illustrated in Fig.~\ref{fig:SmallTau} \ac{w.r.t.} the secrecy rate $R_s\!\leq\!R_t$ and a low rate $R_t$. The accuracy of equation~\eqref{eq:SmallTau0} is illustrated in Fig.~\ref{fig:SmallTau0} \ac{w.r.t.} the rate $R_t$. In both figures we see that the performance in the bulk is superior to the corner. 
\begin{figure*}[!t]
 \centering
  \subfloat[$R_s\!\geq\! 0$]{\includegraphics[width=3.0in]{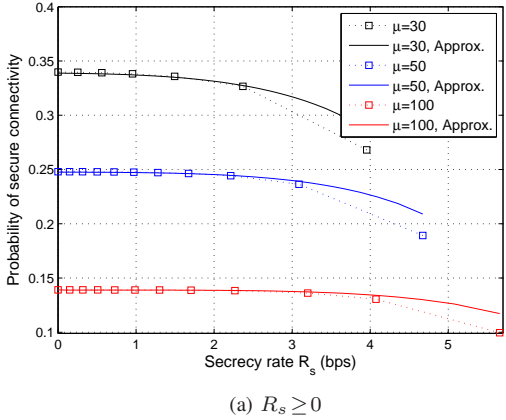} 
\label{fig:HighTau}}
\hfil
  \subfloat[$R_s\!=\! 0$]{\includegraphics[width=3.0in]{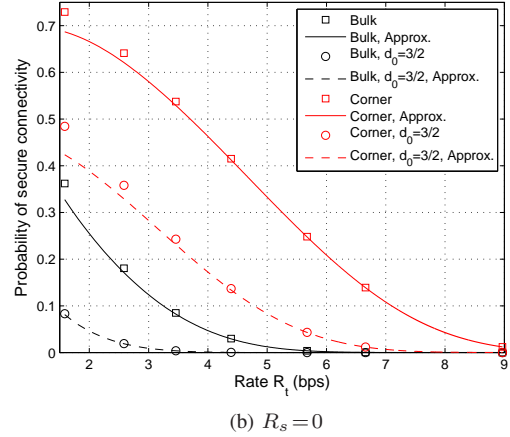} 
\label{fig:HighTau0}}
 \caption{Illustrating the approximation accuracy of the bounds given in equation~\eqref{eq:LargeuCorner} for the corner, and in equation~\eqref{eq:LargeuBulk} for the bulk at high transmission rates $R_t$, low transmission rates $R_s$ and large distance separation $u$. The distance is selected $u\!=\!3$. The rest of the parameter settings can be found in the caption of Fig.~\ref{fig:NewSmallTau}. The exact probability is calculated numerically after substituting equation~\eqref{eq:A} into~\eqref{eq:Psc}. In (a) we depict the results only pertinent to the corner because the probability of secure connectivity at high transmission rates $R_t$ in the bulk is very low. $d_0\!=\!1$ unless otherwise stated.}
\label{fig:NewHighTau}
\end{figure*}
\begin{lemma}
\label{lemma:5}
For high transmission rates $R_t$ and large distance separation $u$ between the receiver and the eavesdropper, the probability of secure connectivity is higher at the corner than in the bulk. 
\begin{proof}
For a large distance separation $u$, we may assume that the interference at the receiver and the eavesdropper is uncorrelated. In that case, the joint connection probability $\mathcal{J}_{\text{x}}\!\left(u\right)$ is equal to the product of the connection probabilities of the receiver and the eavesdropper, and the probability of secure connectivity in equation \eqref{eq:Psc} is simplified to $\mathbb{P}_{\text{x}}^{\text{sc}}\!\left(u\right) \!=\! \mathbb{P}_{\text{x,r}}^{\text{c}}\left(1 -\mathbb{E}\left\{e^{-s_e\mathcal{I}_{\text{x,e}}\left(u\right)}\right\} \right)$. 
For a high transmission rate $R_t$ or equivalently for a large  $\mu$, the connection probability of the receiver at the corner can be approximated by expanding equation~\eqref{eq:Pcb} around $s\!\rightarrow\!\infty$,  
$\mathbb{P}_{\text{co,r}}^{\text{c}}\!\approx\! e^{-\frac{\lambda \pi s^{2/\eta}}{4 \text{sinc}\left( 2\pi / \eta\right) }}$. 
In the bulk, the exponent should be scaled by four, $\mathbb{P}_{\text{bu,r}}^{\text{c}}\!\approx\!e^{-\frac{\lambda \pi s^{2/\eta}}{\text{sinc}\left( 2\pi/\eta \right)}}$. 

The connection probability of the eavesdropper at the boundary, $\mathbb{P}_{\text{co,e}}^{\text{c}}\!\left(u\right) \!=\! \mathbb{E}\!\left\{e^{-s_e\mathcal{I}_{\text{co,e}}\left(u\right)}\right\}$, can be approximated after substituting $I_u$ from equation~\eqref{eq:Iub} 
 into~\eqref{eq:Pc1b}. For a low secrecy rate $R_s$, or equivalently for a large $\sigma$, we can approximate the connection probability of the eavesdropper as $\mathbb{P}_{\text{co,e}}^{\text{c}}\!\approx\! \mathbb{E}_Z\!\left\{ e^{-\frac{\lambda\pi\sigma^{2/\eta}z^{-2/\eta}}{4 \text{sinc}\left( 2\pi/\eta \right)}-\frac{u\lambda\sigma^{1/\eta}z^{-1/\eta}}{\text{sinc}\left(\pi/\eta \right)}}\right\}$. In order to obtain a lower bound for the  probability of secure connectivity at the corner, we can upper-bound the connection probability of the eavesdropper at the boundary. One way to do that is to fix the signal level over the eavesdropper channel at the maximum value $z_2\!=\!\left|u-d_0\right|^{-\eta}$, see the Appendix. Finally, we get  
\begin{equation}
\label{eq:LargeuCorner}
\mathbb{P}_{\text{co}}^{\text{sc}}\!\left(u\right) \gtrsim e^{-\!\frac{\lambda \pi s^{2/\eta}}{4 \text{sinc}\left( 2\pi / \eta\right) }}\left(\!1 \!-\! e^{-\!\frac{\lambda\pi\sigma^{2/\eta}\left(u\!-\!d_0\right)^2}{4 \text{sinc}\left( 2\pi/\eta \right)}} e^{-\!\frac{\lambda\sigma^{1/\eta}u\left(u\!-\!d_0\right)}{\text{sinc}\left(\pi/\eta \right)}} \!\right)\!\!,
\end{equation}
where it is reasonable to assume that $u\!>\!d_0$. 
 
In the bulk, the connection probability of the eavesdropper for a low secrecy rate $R_s$ can be approximated as  $\mathbb{P}_{\text{bu,e}}^{\text{c}}\!\approx\! \mathbb{E}_Z\!\left\{ e^{-\frac{\lambda\pi\sigma^{2/\eta}z^{-2/\eta}}{\text{sinc}\left( 2\pi/\eta \right)}}\right\}$. An upper-bound for the probability of secure connectivity can be obtained by fixing the signal level over the eavesdropper channel at the minimum value $z_1\!=\!\left(d_0^2+u^2\right)^{-\eta/2}$. Finally, we get 
\begin{equation}
\label{eq:LargeuBulk}
\mathbb{P}_{\text{bu}}^{\text{sc}}\!\left(u\right) \lesssim e^{-\frac{\lambda \pi s^{2/\eta}}{\text{sinc}\left( 2\pi / \eta\right) }}\left(1 - e^{-\frac{\lambda\pi\sigma^{2/\eta}\left( d_0^2+u^2\right)}{\text{sinc}\left( 2\pi/\eta \right)}} \right).
\end{equation}

Let us denote $x\!=\!\frac{\lambda \pi}{4 \text{sinc}\left( 2\pi / \eta\right) }$ and $y\!=\!\frac{\lambda u\left( u-d_0\right)}{\text{sinc}\left(\pi/\eta \right)}$. In order to show that  $\lim_{\mu\!\rightarrow\!\infty} \frac{\mathbb{P}_{\text{bu}}^{\text{sc}}\!\left(u\right)}{\mathbb{P}_{\text{co}}^{\text{sc}}\!\left(u\right)}\!<\!1$, it suffices to show that  $\lim_{s\rightarrow\!\infty} \frac{\exp\!\left(-4xs^{2/\eta}\right)\left(1-\exp\!\left(-4x\sigma^{2/\eta}\left(d_0^2+u^2\right) \right)\right)}{\exp\!\left(-xs^{2/\eta}\right)\left(1-\exp\!\left(-x\sigma^{2/\eta}\left(u-d_0\right)^2-y\sigma^{1/\eta}\right)\right)}\!=\!0$, which is true. For $R_s\!=\! 0$, or equivalently $\mu\!=\!\sigma\!\triangleq\!\gamma$, we also get that  $\lim_{\gamma\rightarrow\!\infty} \frac{\exp\!\left(-\frac{4x}{g\left(d_0\right)^{2/\eta}}\gamma^{2/\eta}\right)\left(1-\exp\!\left(-4x\gamma^{2/\eta}\left(d_0^2+u^2\right)\right)\right)}{\exp\!\left(-\frac{x}{g\left(d_0\right)^{2/\eta}}\gamma^{2/\eta}\right)\left(1-\exp\!\left(-x\gamma^{2/\eta}\left(u-d_0\right)^2-y\gamma^{1/\eta}\right)\right)}\!=\!0$. 

When the secrecy rate $R_s$ is high, or equivalently $\sigma$ is low, one can approximate the connection probability of the eavesdropper after substituting equation~\eqref{eq:Iub} into~\eqref{eq:Pc1b} and expanding around $\sigma\!=\! 0$. 
\[ 
\mathbb{P}_{\!\!\text{co,e}}^{\text{c}}\!\left(u\right)\!\approx\! 1 \!-\! \lambda\sigma^{2/\eta}\mathbb{E}_Z\!\!\left\{\!\!\frac{\left(\eta\!-\!1\right)\eta\pi \!\!-\!\! \left(4\!-\!\pi\!+\!\left(\pi\!-\!2\right)\eta\right)u^{2\!-\!\eta}} {2\left(\eta\!-\!1\right)\left(\eta\!-\!2\right)}z^{-\!\frac{2}{\eta}}\!\!\right\}\!.
\]

Since $\sigma\!\rightarrow\! 0$, it is straightforward to show that $\lim\limits_{\mu\!\rightarrow\!\infty}\frac{\mathbb{P}_{\text{bu}}^{\text{sc}}\!\left(u\right)}{\mathbb{P}_{\text{co}}^{\text{sc}}\!\left(u\right)}\!<\!1$, and the proof is complete. 
\end{proof}
\end{lemma}

The intuitive explanation of Lemma~\ref{lemma:5} is as follows: For a large distance separation $u$, the signal level over the eavesdropper channel becomes low, and the probability of secure connectivity at high transmission rates $R_t$ is dominated by the connection probability of the receiver. Therefore the performance is better at the corner, where the interference level is lower than in the bulk. 

The accuracy of the approximations for the probability of secure connectivity  in Lemma~\ref{lemma:5} is illustrated in Fig.~\ref{fig:NewHighTau} for distance separation $u\!=\!3$. At this distance, the correlation coefficient is less than $10^{-1}$ both at the corner and in the bulk, see Fig.~\ref{fig:rho}. In Fig.~\ref{fig:HighTau}, we see that for decreasing $\sigma$, or equivalently, for increasing secrecy rate $R_s$ the approximation accuracy degrades. As expexted, the approximation accuracy improves for increasing rate $R_t$ (or $\mu$). In Fig.~\ref{fig:HighTau}, we also see that for high transmission rates $R_t$, we can allow for increasing secrecy rates $R_s$ over some range, without a noticeable decrease in the probability of secure connectivity. Finally, we note that the approximation given in equation~\eqref{eq:LargeuBulk} for the bulk is an upper bound only for high $\mu,\sigma$ (not visible in Fig.~\ref{fig:HighTau0}).  

For small distances $u$, the impact of correlated interference at the receiver and the eavesdropper should not be ignored. Extending Lemma~\ref{lemma:5} for small $u$ and a positive secrecy rate $R_s$ is tedious. We show the extension only for secrecy rate $R_s\!=\! 0$, or $\mu\!=\!\sigma\triangleq\!\gamma$. For presentation clarity, we will also assume $d_0\!=\!1$. These assumptions are discussed after the proof of this and the following lemma.
\begin{lemma}
\label{lemma:6}
For high transmission rates $R_t$ and small distance separation $u$ between the receiver and the eavesdropper, the probability of secure connectivity is higher at the corner than in the bulk for secrecy rate $R_s\!=\! 0$.  
\begin{proof}
Let us assume that $u\!\leq\! 1$ since the correlation coefficient of interference is large for small distances $u$. In addition, let us assume that $\gamma\!\geq\! 1$ since we consider high transmission rates $R_t$. First, we will approximate the term $\mathcal{J}_{\text{x}}\!\left(u\right)$ at high transmission rates $R_t$, then the connection probability of the receiver. 

In order to approximate the term $\mathcal{J}_{\text{x}}\!\left(u\right)$, we note that for $u\!\leq\! 1$, the signal level over the eavesdropper channel becomes equal to one with probability $p$, while it takes values from the continuous distribution $f_{Z_c}$ with probability $\left(1\!-\! p\right)$, see the Appendix for the definition and the derivation of the \ac{PDF} $f_{Z_c}\!\left(z\right)$. Due to the fact that the \ac{RV} $Z$ follows a mixture distribution for $u\!\leq\! 1$, the quantity $\mathcal{J}_{\text{x}}\!\left(u\right)$ in equation~\eqref{eq:A} can be read as 
\begin{eqnarray}
\label{eq:Au}
\mathcal{J}_{\text{x}}\!\left(u\right) \!\!\!&=&\!\!\! p\,  \exp\left(\!-\! \lambda \! \int_{S_{\text{x}}}\!\! \left(1 \!-\! \frac{1}{1 \!+ \! \gamma  g\!\left(r\right)} \frac{1}{1 \!+\! \gamma g\!\left(d\right)}\right) \! {\rm d} S\right)   + \IEEEnonumber* \\ 
& & \!\!\!\!\!\!\!\!\!\!\!\!\!\!\!\!\!\!\!\!\!\!\!\!\!\!\!\! \left(1\!-\!p\right)\!\!\!\int_{z_1}^{z_2} \!\!\!\!\!\!\! \exp\!\!\left(\!\!-\! \lambda\!\!\! \int\limits_{S_{\text{x}}} \!\!\!\left(\!1 \!-\! \frac{1}{1 \!+ \! \gamma  g\!\left(r\right)} \frac{1}{1 \!+\! \gamma z^{-1} g\!\left(d\right)}\right) \!\! {\rm d} S \!\!\right)\!\!f_{Z_c}\! {\rm d}z 
\end{eqnarray}
where $z_1\!=\!\left(1+u^2\right)^{-\eta/2}$ is the minimum signal level and $z_2$ is the maximum signal level over the eavesdropper channel. 

Next, we show how to approximate the integral $I\!\left(u\right)\!=\! \int_0^{\pi/2}\!\!\! \int_0^{\infty} \!\!\left(1 \!-\! \frac{1}{1 + \gamma  g\left(r\right)} \frac{1}{1 + \gamma g\left(d\right)} \right) {\rm d}S$ at the corner for a large $\gamma$. In order to do that, we will divide  the quadrant $S_{\text{co}}$ into disjoint regions and select a suitable expansion for the terms $\frac{1}{1+\gamma g\!\left(r\right)}$ and $\frac{1}{1+\gamma g\!\left(d\right)}$ over each region. For $\gamma g\!\left(r\right)>1$, or equivalently for $r\!<\!r_0$, where $r_0\!=\!\gamma^{1/\eta}$, we get $\frac{1}{1+\gamma g\!\left(r\right)}\!\approx\!\frac{1}{\gamma g\!\left(r\right)}$. On the other hand, for $r\!>\!r_0$ the term $\gamma g\!\left(r\right)$ becomes small, and $\frac{1}{1+\gamma g\!\left(r\right)}\!\approx\! 1\!-\!\gamma g\!\left(r\right)\!+\!\gamma^2g^2\!\left(r\right)$. The expansion of the term  $\frac{1}{1+\gamma g\!\left(d\right)}$ is more involved as it does not depend only on $r$, but also on $\phi$. We will split the plane into three regions \ac{w.r.t.} the distance $r$: $r\!<\!r_0, r_0\!\leq\! r\!\leq\!r_0\!+\!u$ and $r\!>\!r_0\!+\!u$. The corresponding integral contributions are denoted by $I_j\!\left(u\right)$,  $I\!\left(u\right)\!=\! \sum\nolimits_{j=1}^3 \! I_j\!\left(u\right)$. 

For $r\!<\! r_0$ and angles $\phi\!<\! \phi\!\left(r\right)$, where $\phi\!\left(r\right)=\min\left\{\frac{\pi}{2},\arccos\left(\frac{r^2+u^2-r_0^2}{2ur}\right)\right\}$, the distance $d$ to the eavesdropper is small, thus $\gamma g\!\left(d\right)\!>\! 1$ and $\frac{1}{1+\gamma g\!\left(d\right)}\!\approx\!\frac{1}{\gamma g\!\left(d\right)}$. On the other hand, for $r\!<\! r_0$ and $\phi\!>\! \phi\!\left(r\right)$, the distance to the eavesdropper becomes large, thus the term $\gamma g\!\left(d\right)$ becomes small, and $\frac{1}{1+\gamma g\!\left(d\right)}\!\approx\! 1\!-\!\gamma g\!\left(d\right) \!+\! \gamma^2g^2\!\left(d\right)$. Finally, the integral $I_1\!\left(u\right)$ can be approximated as 
\begin{equation}
\label{eq:A1}
\begin{array}{ccl}
I_1\!\left(u\right) \!\!\! &\gtrsim & \!\!\! \displaystyle    \int_0^{r_0}\!\!\!\!\!\int_0^{\phi\left(r\right)}\!\!\!\left(\!1\!-\!\frac{1}{\gamma^2 \!g\!\left(r\right)g\!\left(d\right)}\!\right) \!{\rm d}\!S + \\ & & \displaystyle  \int_0^{r_0}\!\!\!\!\!\int_{\phi\left(r\right)}^{\frac{\pi}{2}}\!\!\! \left(\!1\!-\!\frac{1-\gamma g\!\left(d\right) + \gamma^2 g^2\!\left(d\right)}{\gamma \!g\!\left(r\right)}\!\right) \!{\rm d}\!S.
\end{array}
\end{equation}

Let us define $r_1\!=\!r_0\!+\!u$. For $r_0 \!\leq r\! \leq \! r_1$, the term $\gamma g\!\left(r\right)$ becomes small, while the term $\gamma g\!\left(d\right)$ might be small or large depending on the angle.  
\begin{equation}
\label{eq:A2}
\begin{array}{ccl}
I_2\!\left(u\right) \!\!\! &\gtrsim & \!\!\! \displaystyle  \int_{r_0}^{r_1}\!\!\!\!\!\int_0^{\phi\left(r\right)}\!\!\!\left(\!1\!-\!\frac{1-\gamma \!g\!\left(r\right)+\gamma^2 \!g^2\!\!\left(r\right)}{\gamma g\!\left(d\right)}\right) \!{\rm d}\!S + \\ & & \displaystyle   \int_{r_0}^{r_1}\!\!\!\!\!\int_{\phi\left(r\right)}^{\frac{\pi}{2}}\!\!\! \left(\!1\!-\!\frac{1-\gamma \!g\!\left(r\right)+\gamma^2 \!g^2\!\!\left(r\right)}{\left(1\!-\!\gamma \!g\!\left(d\right)\!+\!\gamma^2 \!g^2\!\!\left(d\right) \right)^{-1}} \!\right) \!{\rm d}\!S.
\end{array}
\end{equation}

Finally, for $r>r_1$, both terms $\gamma g\!\left(r\right), \gamma g\!\left(d\right)$ become small independent of the angle $\phi$, thus 
\begin{equation}
\label{eq:A3}
I_3\!\left(u\right) \!\gtrsim \! \int_{r_1}^{\infty}\!\!\!\!\!\int_0^{\frac{\pi}{2}}\!\!\! \left(\!1\!-\!\frac{1-\gamma \!g\!\left(r\right)+\gamma^2 \!g^2\!\!\left(r\right)}{\left(1-\gamma \!g\!\left(d\right)+\gamma^2 \!g^2\!\!\left(d\right) \right)^{-1}} \!\right) \!{\rm d}\!S.
\end{equation}

In order to approximate $I\!\left(u\right)$ we need to sum up the approximations from the different regions.  For a small $u$ and a large  $\gamma$, we get that $\phi\!\left(r\right)\!\approx\!\frac{\pi}{2}\, \forall r\!\leq\! r_0$ and $r_1\!\approx\! r_0$. Therefore, the terms that dominate the integral $I\!\left(u\right)$ is the first term in equation~\eqref{eq:A1} and equation~\eqref{eq:A3}. Next, we show how to approximate the dominant terms at a high rate $R_t$ or equivalently for a large $\gamma$. 

Since $r_1\!>\! 1$, the leading order approximation for the integral $\int_0^{\pi/2}\!\!\int_{r_1}^\infty \! \left(\gamma g\!\left(r\right)-\gamma^2g^2\!\left(r\right)\right){\rm d}S$ is calculated  after substituting $g\!\left(r\right)\!=\! r^{-\eta}$ and performing the integration. Finally, we get  $\frac{\eta\pi\gamma^{2/\eta}}{4\left(\eta-1\right)\left(\eta-2\right)}$. On the other hand, the integral $\int_0^{\pi/2}\!\!\int_{r_1}^\infty \!  \left(\gamma g\!\left(d\right)\!-\!\gamma^2\!g^2\!\left(d\right)\right){\rm d}S$ cannot be conputed in closed-form. Nevertheless, after approximating $g\!\left(d\right), g^2\!\left(d\right)$ for a large $r$, i.e.,  $g\!\left(d\right)\!\approx\!r^{-\eta}\!+\!\eta r^{-1-\eta}u\cos\phi$ and $g^2\!\!\left(d\right)\!\approx\!r^{-2\eta}\!+\!2\eta r^{-1-2\eta}u\cos\phi$, which should be valid for $r_1\gg u$, we get  $\left(\frac{\eta\pi\gamma^{2/\eta}}{4\left(\eta-1\right)\left(\eta-2\right)} + \frac{\eta u \gamma^{1/\eta}}{\left(2\eta-1\right)\left(\eta-1\right)}\right)$. Using the large $r$ approximation for $g\!\left(d\right)$, the integral $\gamma^2\int_0^{\pi/2}\!\!\int_{r_1}^\infty \!g\!\left(r\right) g\!\left(d\right)\! {\rm d}S$ is approximated as  $\left(\frac{\pi\gamma^{2/\eta}}{4\left(\eta-1\right)} \!-\! \frac{\left( 2\eta\left(\pi\!-\!1\right) \!-\! \pi\right) u \gamma^{1/\eta}}{4\eta-2}\right)$ for a large $\gamma$. The term $\int_0^{\pi/2}\!\!\!\int_{r_1}^\infty \!\!\left( \!\gamma^3\! \!\left( g\!\left(r\right)\!g^2\!\!\left(d\right) \!+\! g^2\!\!\left(r\right)\!g\!\left(d\right)\right) \!-\! \gamma^4\!g^2\!\!\left(r\right) g^2\!\!\left(d\right)\! \right)\!{\rm d}S$ gives $\left(\frac{\left(5\eta-2\right)\pi\gamma^{2/\eta}}{4\left(2+\eta\left(6\eta-7\right)\right)} - \frac{2\eta+12\eta^2\left(\pi-1\right)-\pi\left(7\eta-1\right)} {2+2\eta\left(12\eta-7\right)}u \gamma^{1/\eta}\right)$, and the term  $\int_0^{\frac{\pi}{2}}\!\!\int_0^{r_0}\!\!\left(\!1\!-\!\frac{1}{\gamma^2 \!g\!\left(r\right)g\!\left(d\right)}\!\right) \!{\rm d}\!S$ gives  $\left(\frac{\eta\pi\gamma^{2/\eta}}{4\left(1+\eta\right)} \!+\! \frac{\eta u \gamma^{1/\eta}}{2\eta+1}\right)$. After summing up, 
\begin{IEEEeqnarray}{ccl}
\label{eq:Aapprox}
I\!\left(u\right) &\gtrsim& \, \frac{\pi}{4}\left( \frac{\eta}{\eta\!+\!1}\!+\!\frac{\eta\!+\!2}{\left(\eta\!-\!1\right) \left(\eta\!-\!2\right)} \!+\! \frac{5\eta\!-\!2} {2\!+\!\eta\left(6\eta\!-\!7\right)}\right)\gamma^{2/\eta} \, + \IEEEnonumber* \\ 
& &  \,\, \Bigg( \frac{\pi}{2} + \frac{3\eta}{\left(4\eta^2\!-\!1\right)\left(\eta\!-\!1\right)} - \\ 
& & \,\,\,\,\,\,\,\,\,\,\,   \frac{2\eta\!+\!12\eta^2\left(\pi\!-\!1\right)\!-\!\pi\left(7\eta\!-\!1\right)} {2\!+\!2\eta\left(12\eta\!-\!7\right)} \Bigg) u \gamma^{1/\eta}. \IEEEyesnumber*
\end{IEEEeqnarray}

In order to approximate the second integral in equation~\eqref{eq:Au},  we note that the behaviour of the term $\gamma z^{-1} g\!\left(d\right)$ depends on the  signal level $z$ which is continuous over $\left[z_1,z_2\right)$. One way to simplify the approximation is to bound the integral using the maximum value of the signal level, 
\begin{equation}
\label{eq:inequality}
\begin{array}{ccc}
\displaystyle \int_{S_{\text{co}}} \!\!\int_{z_1}^{z_2}\!\!\!\left(\!1 \!-\! \frac{1}{1 \!+ \! \gamma  g\!\left(r\right)} \frac{1}{1 \!+\! \gamma z^{-1} g\!\left(d\right)}\!\right)\!\!f_{Z_c}\!\left(z\right) \!{\rm d}z {\rm d} S \!\!\!&\geq&\!\!\! \\ 
\displaystyle \int_{S_{\text{co}}} \!\! \left(1 \!-\! \frac{1}{1 \!+ \! \gamma  g\left(r\right)} \frac{z_2}{z_2\!+\! \gamma g\left(d\right)}\right)  \!\! {\rm d} S. & &  \! 
\end{array}
\end{equation}

We can expand the right-hand side of the above inequality similar to equations~\eqref{eq:A1}$-$\eqref{eq:A3}. Though, we will have to modify some of the integration limits and the way that the angle $\phi\left(r\right)$, separating between small and large distances $d$, is computed. Firstly, in order to calculate the term $I_1\!\left(u\right)$, we will still carry out the integration over $0\!\leq\! r\! \leq \! \gamma^{1/\eta}$, but the angle  $\phi\!\left(r\right)=\min\left\{\frac{\pi}{2},\arccos\left(\frac{r^2+u^2-(\gamma/z_2)^{2/\eta}}{2ur}\right)\right\}$. Secondly, the term $I_2\!\left(u\right)$ is calculated after integrating over $\gamma^{1/\eta} \! \leq \! r \! \leq \! u\!+\! \left(\frac{\gamma}{z_2}\right)^{\! 2/\eta}$, and using the updated expression for $\phi\left(r\right)$. Finally, the term $I_3\!\left(u\right)$ is calculated after integrating over $r\!\geq \! u\!+\! \left(\frac{\gamma}{z_2}\right)^{\! 2/\eta}$. 
\begin{figure}[!t]
 \centering
  \includegraphics[width=3.2in]{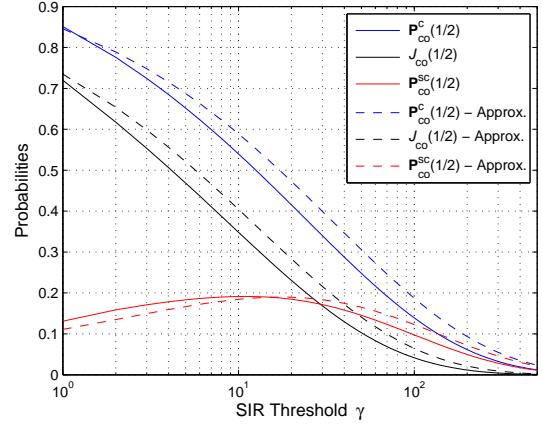}
 \caption{Illustrating the accuracy of the approximations made in Lemma~\ref{lemma:6}. The connection probability of the receiver (blue curves) is calculated  at the corner using equation~\eqref{eq:Pcb} and approximated using equation~\eqref{eq:Pcapprox}. The quantity $\mathcal{J}_{\text{co}}\!\left(u\right)$ (black curves) is calculated numerically based on equation~\eqref{eq:A} and approximated using  equation~\eqref{eq:Aapprox2}. The probability of secure connectivity (red curves) is evaluated after substracting $\mathcal{J}_{\text{co}}$ from the connection probability. Pathloss exponent $\eta\!=\!4$, user density $\lambda\!=\!0.2$, distance $u\!=\!\frac{1}{2}$, $d_0\!=\!1$.}
 \label{fig:Lemma2}
\end{figure}

Note that for $u\!\leq \!1$, we have $z_2\!=\! 1$, and the probability $p$ to experience signal level $Z\!=\! 1$ at the eavesdropper is larger than $\frac{1}{2}$, see the Appendix. Therefore  using $p\!=\! 1$,  in the calculation of $\mathcal{J}_{\text{co}}\!\left(u\right)$ for small distance separations $u\!\leq\! 1$ introduces small  approximation error, see Fig.~\ref{fig:Lemma2} (set of black curves). 
\begin{IEEEeqnarray}{ccl}
\mathcal{J}_{\text{co}}\!\left(u\right) &\gtrsim&  \exp\!\Bigg( \!-\! \frac{\lambda\pi \gamma^{\frac{2}{\eta}}}{4} \!\Big(\frac{\eta}{\eta\!+\!1}\!+\!\frac{\eta\!+\!2} {\left(\eta\!-\!1\right)\left(\eta\!-\!2\right)} + \IEEEnonumber* \\ & &  \frac{5\eta\!-\!2}{2\!+\!\eta\left(6\eta\!-\!7\right)}\Big)\!-  \! \Big( \frac{\pi}{2} \!+\! \frac{3\eta}{\left(4\eta^2\!-\!1\right)\left(\eta\!-\!1\right)} \, - \\ & &   \frac{2\eta\!+\!12\eta^2\left(\pi\!-\!1\right)\!-\!\pi\left(7\eta\!-\!1\right)} {2\!+\!2\eta\left(12\eta\!-\!7\right)} \Big) \!\lambda u \gamma^{1/\eta}\Bigg). \IEEEyesnumber* \label{eq:Aapprox2}
\end{IEEEeqnarray} 

Finally, recall that all approximations made, i.e., expansion for the terms $\frac{1}{1+\gamma g\!\left(r\right)}, \frac{1}{1+\gamma g\!\left(d\right)}$ in equations~\eqref{eq:A1}$-$\eqref{eq:A3}, leading order terms in equation~\eqref{eq:Aapprox} and inequality~\eqref{eq:inequality} are lower bounds to the integrals, thus the approximation would upper-bound the quantity $\mathcal{J}_{\text{co}}\!\left(u\right)$, see Fig.~\ref{fig:Lemma2}. 
\begin{figure*}[!t]
 \centering
  \subfloat[$R_s\!\geq\! 0$]{\includegraphics[width=3.1in]{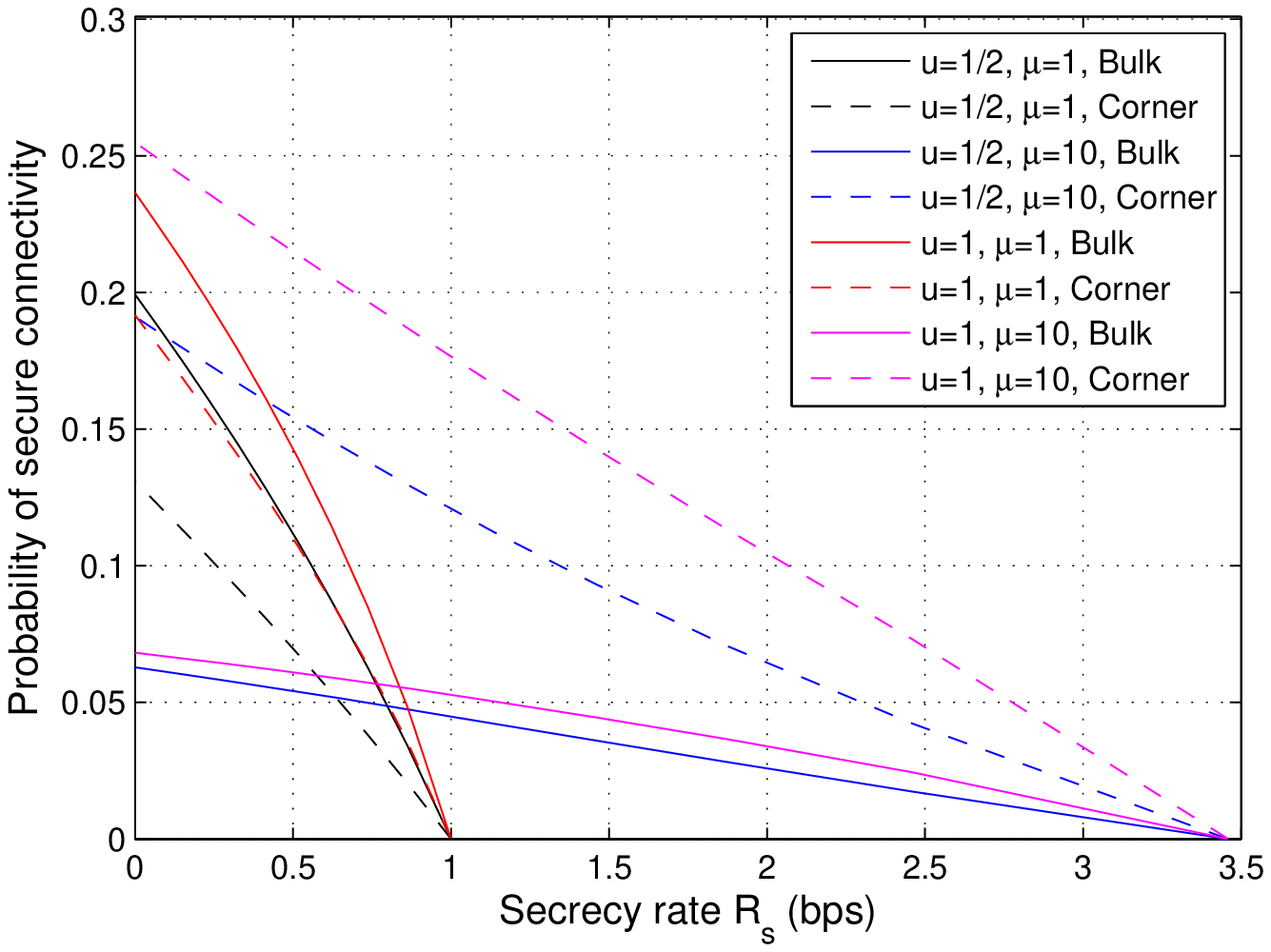} 
\label{fig:Ps}}
\hfil
  \subfloat[$R_s\!=\! 0$]{\includegraphics[width=3.0in]{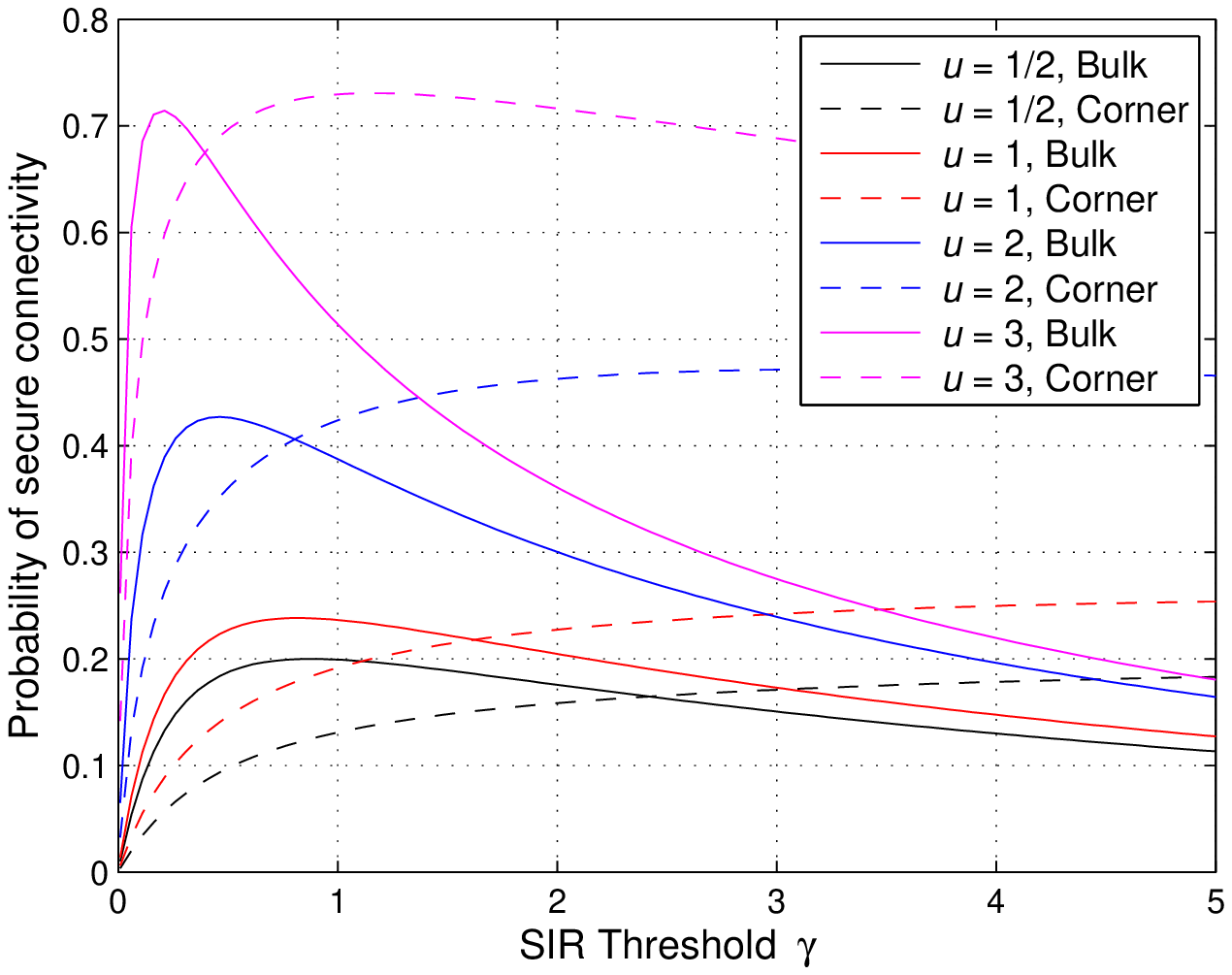} 
\label{fig:Ps0}}
 \caption{Probability of secure connectivity at the corner and in the bulk using equation~\eqref{eq:Psc}. Pathloss exponent $\eta\!=\!4$, user density $\lambda\!=\!0.2$, $d_0\!=\!1$.}
\label{fig:NewPs}
\end{figure*}

Following similar steps, the integral $I\!\left(u\right)$ in the bulk is dominated by the following two terms for a large $\gamma$ and a small distance separation $u$ 
\begin{equation}
\label{eq:Abulk}
\begin{array}{ccl}
I\!\left(u\right) \!\!\! &\gtrsim& \!\!\! \displaystyle 2\!\int_0^{r_0}\!\!\!\!\!\int_0^{\pi}\!\!\!\left(\!1\!-\!\frac{1}{\gamma^2 \!g\!\left(r\right)g\!\left(d\right)}\!\right) \!{\rm d}\!S  +  \\ & & \displaystyle 2\! \int_{r_1}^{\infty}\!\!\!\!\!\int_0^\pi\!\!\! \left(\!1\!-\!\frac{1\!-\!\gamma g\!\left(r\right)\!+\!\gamma^2 \!g^2\!\!\left(r\right)}{\left(1\!-\!\gamma g\!\left(d\right)\!+\!\gamma^2 \!g^2\!\!\left(d\right) \right)^{-1}} \!\right) \!{\rm d}\!S, 
\end{array}
\end{equation}
where the factor $2$ is used to account for angles $\pi\!\leq\!\phi\!\leq\!2\pi$. 

The leading order terms in equation~\eqref{eq:Abulk} can be computed following similar steps to equation~\eqref{eq:Aapprox}
\begin{equation}
\label{eq:Abulk2}
\mathcal{J}_{\!\text{bu}}\!\left(u\right)  \gtrsim  \exp\!\!\left(\!-\!\left(\! \frac{\eta}{\eta\!+\!1}\!+\!\frac{\eta\!+\!2}{\left(\eta\!-\!1\right)\left(\eta\!-\!2\right)} \!+\! \frac{5\eta\!-\!2}{2\!\!+\!\eta\left(6\eta\!-\!7\right)}\!\right)\!\lambda\pi \gamma^{\!\frac{2}{\eta}}\!\right)\!. 
\end{equation}

Comparing with equation~\eqref{eq:Aapprox}, we see that the coefficient of $\gamma^{2/\eta}$ in the bulk, as expected, it is equal to the respective coefficient at the corner after scaling by four. In addition, in the bulk, the approximation of the integral $I\!\left(u\right)$ at high transmission rates does not require a correction term depending on the distance separation $u$. This is due to the following reasons: (i) in the bulk, the mean and the variance of interference are location-independent and, (ii) the terms $2\gamma^2\!\int_0^\pi\!\!\int_{r_1}^\infty\!g\!\left(r\right)\!g\!\left(d\right)\!{\rm d}S$ and $2\gamma^4\!\int_0^\pi\!\!\int_{r_1}^{\infty}\!g^2\!\!\left(r\right)\!g^2\!\!\left(d\right)\!{\rm d}S$ accept a total correction $-4\pi u\gamma^{1/\eta}$ at high rates, but this is cancelled out due to the terms $2\gamma^3\!\int_0^\pi\!\!\int_{r_1}^\infty\!g^2\!\!\left(r\right)\!g\!\left(d\right)\!{\rm d}S$ and  $2\gamma^3\!\int_0^\pi\!\!\int_{r_1}^\infty\!g\!\left(r\right)\!g^2\!\!\left(d\right)\!{\rm d}S$.  

Having approximated the terms $\mathcal{J}_{\text{x}}\!\left(u\right)$, the approximation of the connection probability of the receiver at high transmission rates $R_t$ is rather trivial. It can be done using different expansions for the term $\frac{\gamma g\!\left(r\right)}{1+\gamma g\!\left(r\right)}$ at small and large distances $r$. The connection probability of the receiver at the corner is 
\begin{equation}
\label{eq:Pcapprox}
\begin{array}{ccl}
\mathbb{P}_{\text{co,r}}^{\text{c}} \!\!\!&\gtrsim&\!\!\! \displaystyle  \exp\!\Bigg(\!-\lambda\!\Bigg( \int_0^{\pi/2}\!\!\!\!\int_0^{\gamma^{1/\eta}}\!\!\!\left(1\!-\!\frac{1}{\gamma g\!\left(r\right)}\right){\rm d} S +  \\ & & \displaystyle 
 \int_0^{\pi/2}\!\!\!\!\int_{\gamma^{1/\eta}}^\infty\!\!\left(\gamma g\!\left(r\right)\!-\!\gamma^2\!g^2\!\!\left(r\right) \right){\rm d}S \Bigg)\! \Bigg) \\ 
\!\!\!&\gtrsim&\!\!\! \displaystyle  \exp\!\left(-\!\frac{\lambda \pi}{4}\!\left( \frac{\eta}{\eta\!+\!2} \!+\! \frac{\eta}{\left(\eta\!-\!1\right)\left(\eta\!-\!2\right)}\right)\gamma^{2/\eta} \right).
\end{array}
\end{equation}

The quality of the above approximation for the connection probability is illustrated in Fig.~\ref{fig:Lemma2}, set of blue curves. 

The connection probability in the bulk for a high transmission rate $R_t$ can be approximated following the same steps with equation~\eqref{eq:Pcapprox}, $\mathbb{P}_{\text{bu,r}}^{\text{c}}\!\gtrsim\! e^{-4 \lambda c_1 \gamma^{2/\eta}}$, where $c_1\!=\!\frac{\pi}{4}\left(\frac{\eta}{\eta+2}\!+\!\frac{\eta}{\left(\eta\!-\!1\right)\left(\eta\!-\!2\right)}\right)$. 
 
The ratio of the probabilities of secure connectivity in  the bulk and at the corner as $\gamma\!\rightarrow\!\infty$ is 
\[ 
\lim_{\gamma\!\rightarrow\!\infty}\frac{\mathbb{P}_{\text{bu}}^{\text{sc}}\!\left(u\right)}{\mathbb{P}_{\text{co}}^{\text{sc}}\!\left(u\right)}=\lim_{\gamma\!\rightarrow\!\infty}\frac{e^{-4 c_1\lambda \gamma^{2/\eta}}-e^{-4c_2\lambda \gamma^{2/\eta}}}{e^{-c_1\lambda \gamma^{2/\eta}}-e^{-c_2\lambda \gamma^{2/\eta}}e^{-c_3u\lambda \gamma^{1/\eta}}}\stackrel{(a)}{=}0, 
\]
where $\!c_3\!>\! 0$ is the coefficient of $\!u\gamma^{1/\eta}\!$ in equation~\eqref{eq:Aapprox}, $c_2$ is the coefficient of $\gamma^{2/\eta}$ in equation~\eqref{eq:Aapprox}, and $\!(a)\!$ follows from $0\!<\! c_1\!<\!c_2 \, \forall \eta\!>\!2$.
\end{proof}
\end{lemma}
\begin{lemma}
\label{lemma:3}
For secrecy rate $R_s\!=\! 0$ and small distance separation $u$ between the receiver and the eavesdropper, the transmission rates $R_t\!=\!\log_2\!\left(1+\gamma\right)$ maximizing the probability of secure connectivity in the bulk and at the corner are related as $\gamma_{\text{bu}}^*\!=\! 2^{-\eta}\gamma_{\text{co}}^*$.
\begin{proof}
Using  the leading order $\gamma^{2/\eta}$ in equation~\eqref{eq:Aapprox} and equation~\eqref{eq:Pcapprox}, the probability of secure connecticity at the corner can be approximated as the difference between two exponentials,  $\mathbb{P}_{\text{co}}^{\text{sc}}\!\approx\!e^{-\lambda c_1\gamma^{2/\eta}}\!-\!e^{-\lambda c_2\gamma^{2/\eta}}$. This kind of function accepts  a maximum at $\gamma_{\text{co}}^*\!=\!\left(\frac{\log\left(c_2/c_1\right)}{\lambda\left(c_2-c_1\right)}\right)^{\!\eta/2}$. For $\eta\!=\! 4$, we get $\gamma_{\text{co}}^*\!=\!\frac{11025 \log\left(54/35\right)}{361 \lambda^2\pi^2}$, which is close to the value seen in Fig.~\ref{fig:Lemma2}. Similarly, the transmission rate maximizing the probability of secure connectivity in the bulk is $\gamma_{\text{bu}}^*\!=\! \left(\frac{\log\left(c_2/c_1\right)}{4\lambda\left(c_2-c_1\right)}\right)^{\!\eta/2}$, thus $\gamma_{\text{bu}}^*\!=\! 2^{-\eta}\gamma_{\text{co}}^*$.
\end{proof}
\end{lemma}

Recall that in Lemma~\ref{lemma:6} the plane has been divided into three areas \ac{w.r.t.} to the distance $r$ from the receiver, i.e., $r\!\leq\!r_0, r_0\!<r\!<r_1$ and $r\!\geq\!r_1$. The extension of Lemma~\ref{lemma:6} for positive secrecy rates $R_s\!>\!0$ is tedious because for $\mu\neq\sigma$ we need to separate between more cases while identifying the areas where the terms $\mu g\!\left(r\right)$ and $\sigma g\!\left(d\right)$ accept different expansions. Apart from that, the proof will follow exactly the same steps with Lemma~\ref{lemma:6}. Furthermore, given $R_s\!=\!0$, a generalization for arbitrary $d_0$ will only modify the separation distance to  $r_0\!=\!\left(\frac{\gamma}{g\left(d_0\right)}\right)^{1/\eta}$ instead of $r_0\!=\!\gamma^{1/\eta}$ used in Lemma~\ref{lemma:6}. The expressions in Lemma~\ref{lemma:6} and Lemma~\ref{lemma:3} will also look more complicated because for $d_0\!>\!1$, $s\!=\!\frac{\gamma}{g\left(d_0\right)}$ instead of $s\!=\!\gamma$.

In Fig.~\ref{fig:Ps} we see that the probability of secure connectivity is higher in the bulk for low transmission rates $R_t$ (corresponding to $\mu\!=\! 1$) confirming Lemma~\ref{lemma:4}, and higher at the corner for high transmission rates (corresponding to $\mu\!=\! 10$), confirming Lemma~\ref{lemma:6}. Same behaviour is observed for $R_s\!=\! 0$ in Fig.~\ref{fig:Ps0}, where we see that for increasing distance separation between the receiver and the eavesdropper, the receiver performance at the corner outweighs the performance in the bulk over a wider range of transmission rates $R_t$. 
\begin{figure*}[!t]
 \centering
  \subfloat[$d_0\!=\! \frac{1}{2}$]{\includegraphics[width=2.2in]{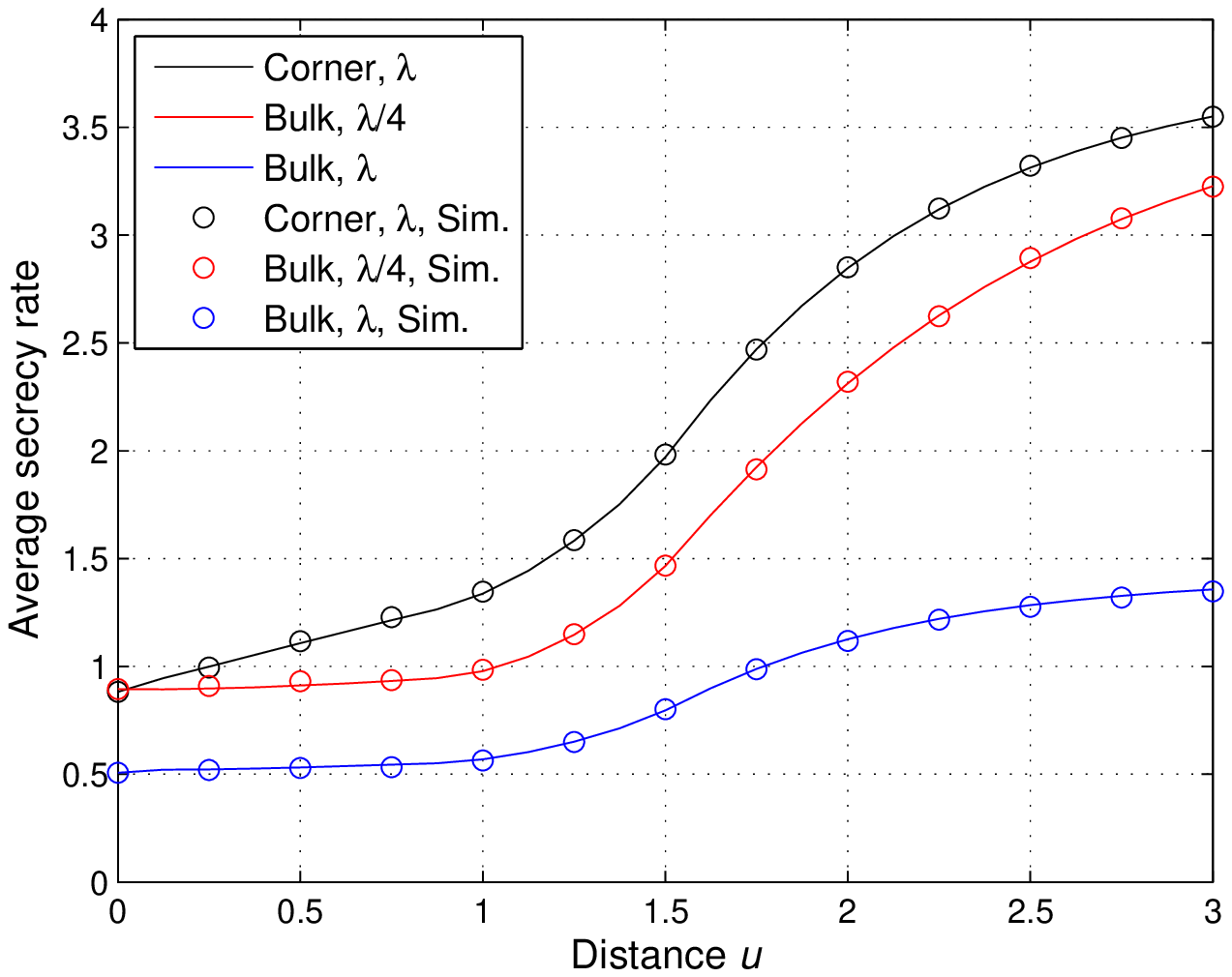} 
\label{fig:Rated0}}
\hfil
  \subfloat[$d_0\!=\! 1$]{\includegraphics[width=2.2in]{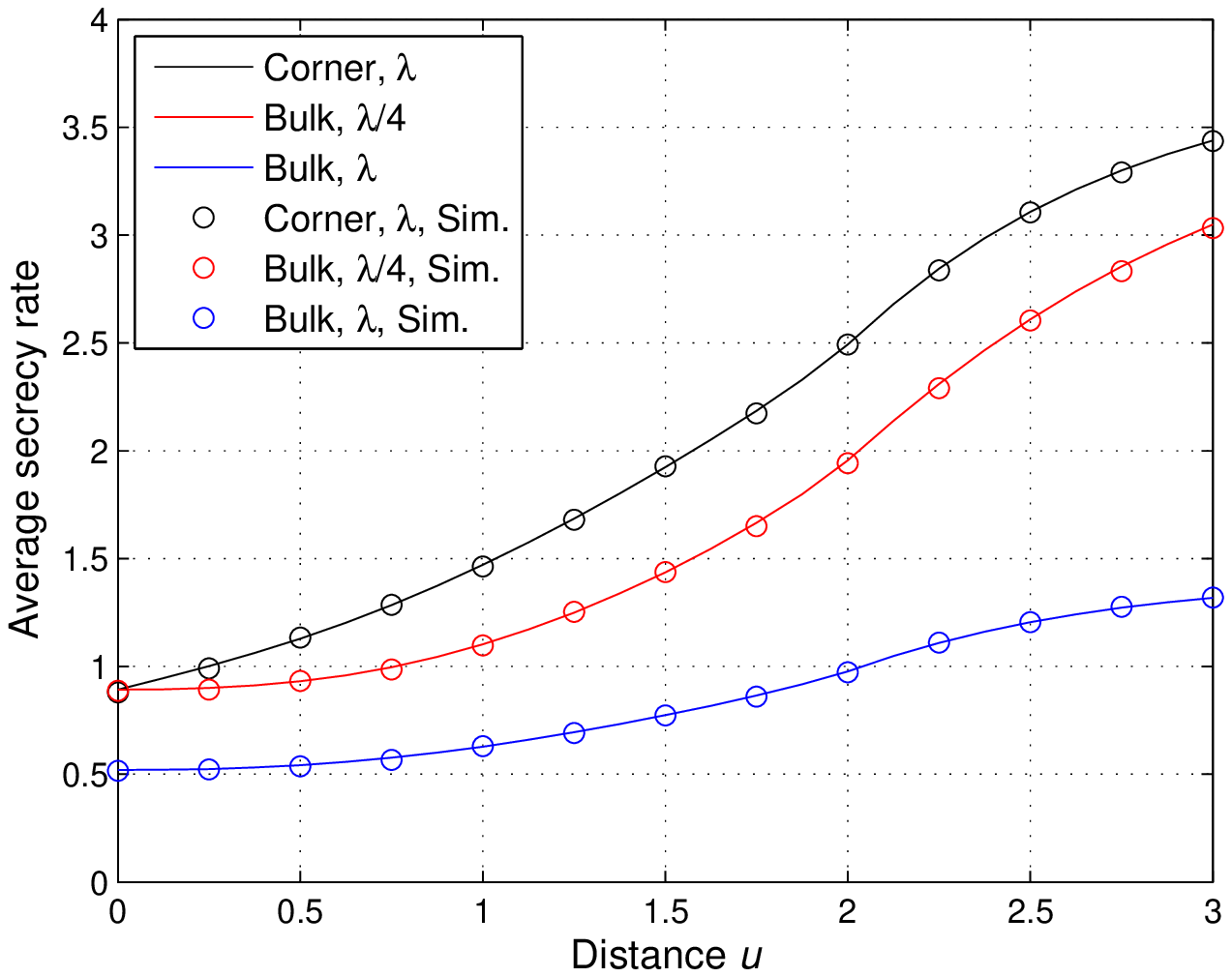} 
\label{fig:Rate}}
\hfil
  \subfloat[$d_0\!=\! \frac{3}{2}$]{\includegraphics[width=2.2in]{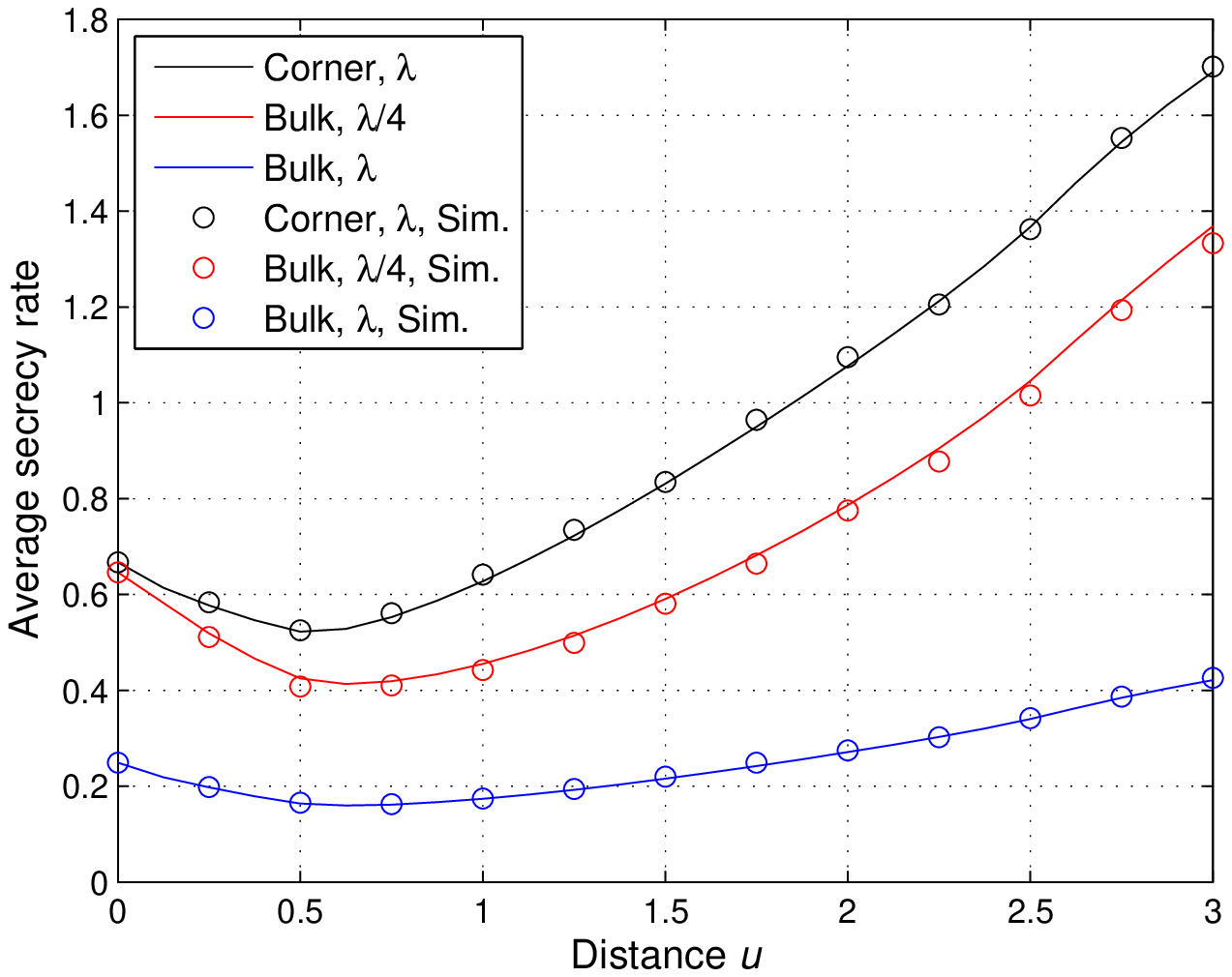} 
\label{fig:Rated0_32}}
 \caption{Average capacity with secrecy assuming known \ac{CSI} at the transmitter \ac{w.r.t.} to the eavesdropper location. In the bulk, the results are generated for intensity of interferers equal to $\lambda$ as well as $\frac{\lambda}{4}$. Pathloss exponent $\eta\!=\!4$, user intensity  $\lambda\!=\!0.2$.}
\label{fig:Rate}
\end{figure*}

\section{Average secrecy capacity $-$ known \ac{CSI}}
\label{sec:Secrecy2}
Let us denote by $f\!\left(\gamma_{\text{r}},\gamma_{\text{e}}\right)$ and  $F\!\left(\gamma_{\text{r}},\gamma_{\text{e}}\right)$ the \ac{PDF} and the \ac{CDF} of the joint \ac{SIR} distribution at the receiver and the eavesdropper, where the dependency on the location is omitted for brevity. The inner integral in equation~\eqref{eq:Rate0} can be read as  
\begin{IEEEeqnarray}{ccl}
I_{\!R} \, & = & \, \int_0^{\gamma_{\text{r}}}\! \left( \log_2\!\left(1+\gamma_{\text{r}}\right) - \log_2\!\left(1+\gamma_{\text{e}}\right) \right)  f\!\left(\gamma_{\text{r}},\gamma_{\text{e}}\right) {\rm d}\gamma_{\text{e}} \IEEEnonumber* \\ 
\, & = & \,  \log_2\!\left(1+\gamma_{\text{r}}\right) \int_0^{\gamma_{\text{r}}} \!\! f\!\left(\gamma_{\text{r}},\gamma_{\text{e}}\right) {\rm d}\gamma_{\text{e}} - \\ & & \,\,\,\,\,  \int_0^{\gamma_{\text{r}}} \!\! \log_2\!\left(1+\gamma_{\text{e}}\right) \frac{\partial^2\! F\!\left(\gamma_{\text{r}},\gamma_{\text{e}}\right)}{\partial\gamma_{\text{r}}\partial\gamma_{\text{e}}} {\rm d}\gamma_{\text{e}} \IEEEnonumber* \\ 
\, &\stackrel{(a)}{=}& \, \log_2\!\left(1+\gamma_{\text{r}}\right) \int_0^{\gamma_{\text{r}}} \!\! f\!\left(\gamma_{\text{r}},\gamma_{\text{e}}\right) {\rm d}\gamma_{\text{e}} \!-\! \left[\log_2\!\left(1\!+\!\gamma_{\text{e}}\right) \frac{\partial F}{\partial \gamma_{\text{r}} }\right]_0^{\gamma_{\text{r}}} \!\!+\! \\ & &  \,\,\,\,\, \frac{1}{\log\!\left(2\right)} \int_0^{\gamma_{\text{r}}}\!\! \frac{1}{1+\gamma_{\text{e}}}\frac{\partial F}{\partial\gamma_{\text{r}}}{\rm d}\gamma_{\text{e}} \IEEEyesnumber\IEEEyessubnumber*  \label{eq:Rate10} \\ 
\, &\stackrel{(b)}{=}& \, \frac{1}{\log\!\left(2\right)} \int_0^{\gamma_{\text{r}}}\! \frac{1}{1+\gamma_{\text{e}}}\frac{\partial F}{\partial\gamma_{\text{r}}}{\rm d}\gamma_{\text{e}}, \label{eq:Rate1}
\end{IEEEeqnarray}
where $(a)$ uses integration by parts, and $(b)$ uses that $\frac{\partial F}{\partial\gamma_{\text{r}}}\!=\!\int_0^{\gamma_{\text{e}}}f\!\left(\gamma_{\text{r}},y\right){\rm d}y$, thus the first two terms in equation~\eqref{eq:Rate10} are  cancelled out.

\noindent After substituting equation~\eqref{eq:Rate1} into equation~\eqref{eq:Rate0} we get 
\begin{IEEEeqnarray}{ccl}
\overline{C}_{\text{x}}^{\text{sc}}\!\!\left(u\right) \, &=& \, \frac{1}{\log\!\left(2\right)} \int_0^\infty\!\!  \int_0^{\gamma_{\text{r}}}\! \frac{1}{1+\gamma_{\text{e}}}\frac{\partial F}{\partial\gamma_{\text{r}}}{\rm d}\gamma_{\text{e}} {\rm d}\gamma_{\text{r}} \IEEEnonumber* \\ &\stackrel{(a)}{=}& \frac{1}{\log\!\left(2\right)} \int_0^\infty\!\!  \int_{\gamma_{\text{e}}}^\infty\! \frac{1}{1+\gamma_{\text{e}}}\frac{\partial F}{\partial\gamma_{\text{r}}}{\rm d}\gamma_{\text{r}} {\rm d}\gamma_{\text{e}}  \\ 
\, & = & \, \frac{1}{\log\!\left(2\right)} \! \int_0^\infty\!\! \frac{1}{1\!+\!\gamma_{\text{e}}} \! \left[ F\!\left(\gamma_{\text{r}},\gamma_{\text{e}}\right) \right]_{\gamma{\text{e}}}^\infty  {\rm d}\gamma_{\text{e}} \\ &\stackrel{(b)}{=}& \frac{1}{\log\!\left(2\right)} \!\int_0^\infty\!\! \frac{1}{1\!+\!\gamma_{\text{e}}} \!\left( F_{{\text{e}}}\!\!\left(\gamma_{\text{e}}\right) \!-\! F\!\left(\gamma_{\text{e}},\gamma_{\text{e}} \right)\right) {\rm d}\gamma_{\text{e}} \IEEEnonumber* \\ \, & = & \, \frac{1}{\log\!\left(2\right)} \!\int_0^\infty\!\! \frac{1}{1\!+\!\gamma} \!\left( F_{{\text{e}}}\!\!\left(\gamma\right) \!-\! F\!\left(\gamma,\gamma \right)\right) {\rm d}\gamma \\ &\stackrel{(c)}{=}& \frac{1}{\log\!\left(2\right)} \!\int_0^\infty\! \frac{\mathbb{E}\!\left\{ e^{-\gamma \mathcal{I}_{\text{x,r}}}\right\} - \mathcal{J}_{\text{x}}\!\left(u,\gamma\right)}{1\!+\!\gamma} {\rm d}\gamma, \IEEEyesnumber* \label{eq:Rate2}
\end{IEEEeqnarray}
where in $(a)$ we have changed the order of integration, $(b)$ uses that $F\left(\infty,\gamma_{\text{e}}\right)\!=\! F_{{\text{e}}}\!\left(\gamma_{\text{e}}\right)$, see for instance~\cite[Chapter 6]{Papoulis1991}, with $F_{{\text{e}}}\!\left(\gamma_{\text{e}}\right)$ being the \ac{CDF} of the \ac{SIR} at the eavesdropper, and $(c)$ uses that  $F_{{\text{e}}}\!\!\left(\gamma\right)\!=\! 1 -\mathbb{E}\!\left\{ e^{-\gamma z^{-1}\mathcal{I}_{\text{x,e}}\left(u\right)}\right\}$, $F\!\left(\gamma,\gamma \right)\!=\! \mathbb{E}\!\left\{ \left(1\!-\! e^{-\gamma z^{-1}\mathcal{I}_{\text{x,e}}\left(u\right)}\right) \left( 1\!-\! e^{-\gamma \mathcal{I}_{\text{x,r}}} \right) \right\}$, and $\mathcal{J}_{\text{x}}\!\left(u,\gamma\right)$ stands for the joint connection probability of the receiver and the eavesdropper given in equation~\eqref{eq:A} for $\mu\!=\!\sigma\!\triangleq\!\gamma$.

Equation~\eqref{eq:Rate2} indicates that for computing the average secrecy capacity with known \ac{CSI} one has to integrate the probability of secure connectivity in equation~\eqref{eq:Psc} for $\mu\!=\!\sigma\!\triangleq\!\gamma$ over the derivative of the rate function. Another way to put equation~\eqref{eq:Rate2} is to see that the transmitter has to sacrifice some of its rate to achieve \ac{PLS}, and the amount of loss depends on the location of the receiver, the eavesdropper and the interference effects, incorporated into the quantity $\mathcal{J}_{\text{x}}\!\left(u,\gamma\right)$.
\begin{equation}
\label{eq:Rate}
\overline{C}_{\text{x}}^{\text{sc}}\!\!\left(u\right) = \overline{C}_{\text{x}} - \frac{1}{\log\!\left(2\right)}\int_0^\infty \frac{\mathcal{J}_{\text{x}}\!\left(u,\gamma\right)}{1+\gamma} {\rm d}\gamma, 
\end{equation}
where the average transmission rate without secrecy is $\overline{C}_{\text{x}}\!=\!\frac{1}{\log\left(2\right)}\int_0^\infty \frac{1-F_{\text{r}}\left(\gamma\right)}{1+\gamma} {\rm d}\gamma$, see for instance~\cite{Lin2013}, and $F_{\text{r}}\!\left(\gamma\right)$ is the \ac{CDF} of the \ac{SIR} at the receiver. 

\begin{figure}[!t]
 \centering
  \includegraphics[width=3.2in]{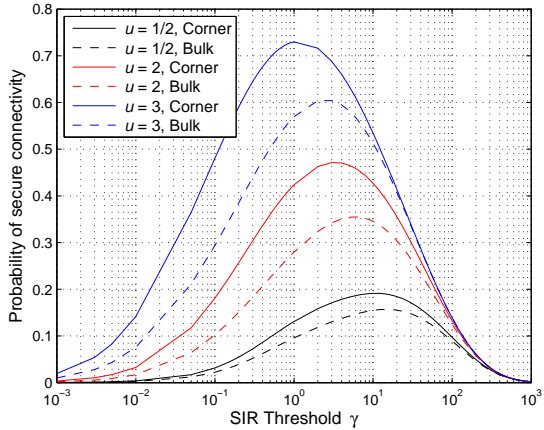}
 \caption{Probability of secure connectivity at the corner and in the bulk \ac{w.r.t.} the SIR threshold $\gamma$ and the location $u$ of the eavesdropper. Pathloss exponent $\eta\!=\!4$. The user density is $\lambda_{\text{co}}\!=\!0.2$ at the corner, and $\lambda_{\text{bu}}\!=\!0.05$ in the bulk.}
 \label{fig:Interplay}
\end{figure}

For independent interference at the receiver and the eavesdropper, $F\!\left(\gamma,\gamma\right) \!=\! F_{\text{e}}\!\!\left(\gamma\right) F_{\text{r}}\!\left(\gamma\right)$, and the analysis in~\cite[Equation (12)]{Wang2014} is confirmed. 
In Fig.~\ref{fig:Rate}, we depict the average capacity with secrecy after evaluating equation~\eqref{eq:Rate} numerically. The results are also verified by simulations. We see that placing the receiver at the corner offers higher average capacity for all distances between the receiver and the eavesdropper, even if the intensity of interferers is four times higher than the intensity of interferers in the bulk. 
\begin{lemma}
\label{lemma:7}
The average secrecy capacity with known \ac{CSI} at the transmitter is higher when the receiver is located at the corner than in the bulk even if the intensity of interferers at the corner is four times higher. 
\begin{proof}
Based on equation~\eqref{eq:Rate2}, it suffices to show that the probability of secure connectivity for $\mu\!=\!\sigma\!\triangleq\!\gamma$ is higher at the corner than in the bulk for all $\gamma$. This is possible to show as follows: Firstly, for a high $\gamma$ and a small distance $u$, the probability of secure connectivity at the corner can be expressed, according to Lemma~\ref{lemma:6}, as  $\mathbb{P}_{\text{co}}^{\text{sc}}\left(\lambda\right) \!\approx\!e^{-\lambda c_1\gamma^{2/\eta}}\!-\!e^{-\lambda c_2\gamma^{2/\eta}} e^{-\lambda c_3 u \gamma^{1/\eta}}$. In the bulk, the probability of secure connectivity is $\mathbb{P}_{\text{bu}}^{\text{sc}}\left(\frac{\lambda}{4}\right)\!\approx\! e^{-4 \frac{\lambda}{4}  c_1\gamma^{2/\eta}}\!-\!e^{-4 \frac{\lambda}{4} c_2\gamma^{2/\eta}}$, see Lemma~\ref{lemma:6}. For a positive $u\!>\!0$,  $\mathbb{P}_{\text{bu}}^{\text{sc}}\!\left(\frac{\lambda}{4}\right)\!<\! \mathbb{P}_{\text{co}}^{\text{sc}}\!\left(\lambda\right)$ due to the fact that $c_3\!>\! 0 \, \forall \eta\!>\!2$. Only in the limit $\gamma\!\rightarrow\!\infty$, we get that  $\mathbb{P}_{\text{co}}^{\text{sc}}\!\left(\lambda\right)\!=\!\mathbb{P}_{\text{bu}}^{\text{sc}}\!\left(\frac{\lambda}{4}\right)$. If the distance separation is large, we can use the approximations in Lemma~\ref{lemma:5}, $\mathbb{P}_{\text{co}}^{\text{sc}}\left(\lambda\right) \!\approx\! e^{-x\gamma^{2/\eta}}\left(1 - \mathbb{E}_Z\!\left\{ e^{-x\gamma^{2/\eta}z^{-2/\eta}-y\gamma^{1/\eta}z^{-1/\eta}}\right\}\right)$ and $\mathbb{P}_{\text{bu}}^{\text{sc}}\left(\frac{\lambda}{4}\right)\!\approx\! e^{-x\gamma^{2/\eta}}\left(1 - \mathbb{E}_Z\!\left\{ e^{-x\gamma^{2/\eta}z^{-2/\eta}}\right\}\right)$, thus $\mathbb{P}_{\text{bu}}^{\text{sc}}\!\left(\frac{\lambda}{4}\right)\!<\! \mathbb{P}_{\text{co}}^{\text{sc}}\!\left(\lambda\right)$ for $y\!>\!0$.  Secondly, for a low $\gamma$, the probability of secure connectivity is dominated by the mean interference at the eavesdropper, see equation~\eqref{eq:SmallTau}. Due to the scaled intensity of users, the mean interference at the receiver is equal at the corner and in the bulk, however, the mean interference at the eavesdropper is higher at the boundary  for $u\!>\!0$ than in the bulk. Therefore $\mathbb{P}_{\text{bu}}^{\text{sc}}\left(\frac{\lambda}{4}\right)\!<\! \mathbb{P}_{\text{co}}^{\text{sc}}\left(\lambda\right)$ for a low $\gamma$ too. Example illustration using numerical integration of equation~\eqref{eq:Psc} is available in Fig.~\ref{fig:Interplay}. 
\end{proof}
\end{lemma}

\section{Conclusions}
\label{sec:Conclusions}
In this paper, we have calculated the probability of secure connectivity and the average secrecy capacity in a Poisson field of interferers. The receiver performance has been assessed in the infinite plane and also at the corner of a quadrant, and the results have been compared. The analysis shows that hiding the receiver at the corner can provide secrecy enhancement for high dara rate applications. Exposing the receiver at less interference than the eavesdropper is beneficial for physical layer security, even if the boundaries enhance the spatial correlation of interference. On the other hand, for low-rate applications, the impact of boundaries is detrimental because the interference level is reduced overall, and the eavesdropper can mostly decode the low-rate transmissions. In that case, applying further secrecy enhancement techniques, e.g., transmission of jamming signals could be of use to increase the interference level at the boundary. Studying the performance of secrecy enhancement techniques over finite areas, also with more complex geometries, is a direction for future work.  

\appendix 
The \ac{RV} $X$ describing the distance between the transmitter and the eavesdropper, $X\!=\! \|d_0e^{j\Theta}-u\|$, ranges in $\left[\left|u-d_0\right|,\sqrt{d_0^2+u^2}\right]$. Due to the fact that the \ac{RV} $\Theta$ follows the uniform \ac{PDF} in $\left[0,\frac{\pi}{2}\right]$, one may calculate the \ac{PDF} of $X$ 
\[ 
f_X\!\left(x\right) \!=\! \frac{4x}{\pi \sqrt{x^2-\left(u-d_0\right)^2} \sqrt{\left(d_0+u\right)^2-x^2}}.
\]

For $u\!\leq\!\sqrt{1-d_0^2}$ with $d_0\!\leq\! 1$, the distance $X$ becomes smaller than one with probability $p\!=\!1$. After integrating the \ac{PDF} of the distance $f_X\!\left(x\right)$, one can calculate that for $\left\{u\!\leq\!d_0,u\!\geq\!d_0-1\right\}$ and $\left\{u\!\geq\!d_0,u\!\leq\!d_0+1\right\}$, the distance becomes smaller than one with probability 
\[
p\!=\! 1-\frac{2}{\pi} \arctan\!\left( \sqrt{ \frac{\left(d_0^2\left(1\!+\!\alpha^2\right)\!-\!1\right)^2}{\left(2\alpha d_0^2\right)^2-\left(d_0^2\left(1\!+\!\alpha^2\right)\!-\!1\right)^2}} \right), 
\]
where $\alpha\!=\!\frac{u}{d_0}$.

Due to the fact that the distance-based pathloss $g\!\left(r\right)$ becomes equal to unity for distances smaller than one, the \ac{RV} $Z$ follows in general the mixture distribution $f_Z\!\left(z\right) = p\,\delta\!\left(z-1\right) + \left(1-p\right) f_{Z_c}\!\left(z\right)$, where  $\delta\left(\cdot\right)$ is the Dirac delta function. The \ac{PDF} of the continuous \ac{RV} $Z_c\!=\!\|d_0 e^{j\Theta}-u\|^{-\eta}\!:\!Z_c\!<\!1$ can be derived from the distance distribution $f_X\!\left(x\right)$ and it is equal to 
\[ 
f_{Z_c}\!\left(z\right) \!=\! \frac{1}{1\!-\!p} \frac{4 z^{-\!1-\!\frac{2}{\eta}} \left(\pi \eta\right)^{-1}}{\sqrt{z^{-\!\frac{2}{\eta}}\!-\! \left(u\!-\!d_0\right)^2} \!\!\sqrt{\left(d_0\!+\!u\right)^2 \!-\! z^{-\!\frac{2}{\eta}}}}, z_1\!\leq\! z \!<\! z_2, 
\]
where $z_1\!=\!\left(d_0^2+u^2\right)^{-\frac{\eta}{2}}$ and $z_2\!=\!1$. 

For the pairs $\left\{u,d_0\right\}$ giving $p\!=\!0$, the \ac{RV} $Z$ becomes continuous in $\left[z_1,z_2\right]$ where $z_2\!=\!\left|u-d_0\right|^{-\eta}$. 

For the simplified case $d_0\!=\!1$, $p\!=\!\frac{4}{\pi}\arctan\!\left(\sqrt{\frac{2-u}{2+u}}\right)$ for $u\!\leq\!2$, and $p\!=\!0$ for $u\!>\!2$. In that case, the computation of the mean link gain can take a compact form for pathloss exponent $\eta\!=\!4$. We give below the expressions for the mean link gain for $u\!\geq\! 2$. For $u\!<\! 2$,  we give the mean of the continuous part of the distribution
\[
\begin{array}{ccl}
\mathbb{E}\left\{Z\right\}  \!\!\!&\!=\!&\!\!\! \frac{4u\left(u^2-1\right) + \left(1+u^2\right)^2\left(\pi+4\arctan\left(\frac{1}{u}\right)\right)}{\pi \left(u^2-1\right)^3 \left(1+u^2\right)},  u\!\geq\! 2. \\ 
\mathbb{E}\left\{Z_c\right\} \!\!\! &\!=\!& \!\!\!  \frac{2\left(1+u^2\right) \left(2\arctan\left(\frac{1}{u}\right) - \arcsin\left(\frac{3u-u^3}{2}\right)\right)}{\left(1\!-\!p\right)\pi\left(u^2\!-\!1\right)^3} + \\ & & \frac{2u\left(2\!-\!\left(1\!+\!u^2\right)\sqrt{4\!-\!u^2} \right)}{\left(1\!-\!p\right)\pi\left(u^2\!-\!1\right)\left(u^4\!-\!1\right)},  1\!<\! u \! < \! 2. \\ 
\mathbb{E}\left\{Z_c\right\} \!\!\! &\!=\!& \!\!\! \frac{3\sqrt{3}-2}{2\pi}, u\!=\! 1. \\ 
\mathbb{E}\left\{Z_c\right\} \!\!\! &\!=\!& \!\!\! \frac{2\left(1+u^2\right) \left(2\arctan\left(\frac{1}{u}\right) + \arcsin\left(\frac{3u-u^3}{2}\right)-\pi \right)}{\left(1\!-\!p\right)\pi\left(u^2\!-\!1\right)^3} + \\ & &  \frac{2u\left(2\!-\!\left(1\!+\!u^2\right)\sqrt{4\!-\!u^2} \right)}{\left(1\!-\!p\right)\pi\left(u^2\!-\!1\right)\left(u^4\!-\!1\right)},  0\!<\! u \! < 1.
\end{array}
\]

\end{document}